\documentclass[journal]{IEEEtran}
\usepackage{cite}
\usepackage{graphicx}
\usepackage{psfrag}
\usepackage{subfigure}
\usepackage{color}

\usepackage{amsmath}
\usepackage{array}
\usepackage{graphics}
\usepackage{epsfig}
\usepackage{amsbsy}
\usepackage{amssymb}
\usepackage{amsthm}
\newtheorem{theorem}{Theorem}
\newtheorem{lemma}[theorem]{Lemma}

\newtheorem{remark}[theorem]{Remark}

\usepackage[keeplastbox]{flushend}
\usepackage{framed}
\usepackage{algorithm}
\newtheorem{proposition}{Proposition} 
\IEEEoverridecommandlockouts

\begin{document}

\title{Interference-Aided Energy Harvesting: Cognitive Relaying with Multiple Primary Transceivers}
\author{Sanket S.~Kalamkar~\IEEEmembership{Member,~IEEE,} and~Adrish~Banerjee~\IEEEmembership{Senior~Member,~IEEE}
\thanks{S. S. Kalamkar is with the Department of Electrical Engineering, University of Notre Dame, 46556, IN, USA (e-mail: skalamka@nd.edu). A. Banerjee is with the Department of Electrical Engineering, IIT Kanpur, Kanpur, 208016, India (e-mail: adrish@iitk.ac.in).}
\thanks{This work was carried out while S. S. Kalamkar was at IIT Kanpur, India.}
\thanks{Part of this work was presented at 2015 IEEE Global Communications Conference (GLOBECOM'15)~\cite{sank_globecom}.}
}

\maketitle

\begin{abstract}
We consider a spectrum sharing scenario where a secondary transmitter (ST) communicates with its destination via a decode-and-forward secondary relay (SR) in the presence of interference from multiple primary transmitters. The SR harvests energy from received radio-frequency signals that include primary interference and uses it to forward the information to the secondary destination. The relay adopts a time switching policy that switches between energy harvesting and information decoding over the time. Under the primary outage constraints and the peak power constraints at both ST and SR, to determine the average secondary throughput, we derive exact analytical expressions for the secondary outage probability and the ergodic capacity, which characterize the delay-limited and the delay-tolerant transmissions, respectively. We also investigate the effects of the number of primary transceivers and the peak power constraints on the optimal energy harvesting time that maximizes the secondary throughput. By utilizing the primary interference as an energy source, the secondary network achieves a better throughput performance compared to the case where the primary interference is ignored for energy harvesting purpose. Finally, we consider a case where ST also harvests energy from primary transmissions and compare its throughput performance with that of the non-energy harvesting ST case.
\end{abstract}

\section{Introduction}
Energy harvesting (EH) cognitive radio~\cite{sultan,lee1,jeya,usman,shaf} is a promising solution to the problem of the inefficient spectrum use while achieving green communications. Cognitive radio can improve the spectral efficiency by facilitating the unlicensed/secondary users (SUs) to share the spectrum with the licensed/primary users (PUs), provided that the interference to PUs stays below a specified threshold. On the other hand, energy harvesting provides the cognitive radio a greener alternative to harness energy for its operation, which also helps enhance its lifetime under the energy constraint.

Besides harvesting energy from natural sources like solar and wind, nowadays, the radio environment can feed energy in the form of radio-frequency (RF) signals~\cite{lu}. Noticing that RF signals can carry both information and energy together, the authors in~\cite{varshney,grover,rui2} advocated the use of RF signals to harvest energy along with the information transmission. But it is difficult for a receiver, in practice, to simultaneously decode information and harvest energy from received RF signals. Hence two practical policies were proposed to harvest energy and decode information separately~\cite{rui2,rui3, nasir}: The first policy is the time switching policy where the time is switched between energy harvesting and information decoding. The second policy is based on power splitting where a part of the received power is used to harvest energy and the rest for the information decoding. 

Such a wireless energy harvesting while receiving information has an important application in cooperative relaying, where an intermediate node helps forward the information from the source to the destination and improves the coverage and the reliability of the communication~\cite{nasir,aissa_TWC,krik,yener1,ishibashi1,poor,gan,micha,nasir2,chen1}. But the relay usually has a battery of limited capacity, which needs to be replaced or recharged frequently. In this case, wireless energy harvesting can help the relay to stay active in the network and facilitate the information cooperation without frequently replacing or recharging the battery.

\subsection{Motivation}
In spectrum sharing, both PU and SU transmit together which limits the transmit powers of the secondary transmitter (ST) and the secondary relay (SR) in order to keep the interference to PU below a threshold. But PU, being a legacy user, has no such restriction on its transmit power. Hence SU may experience heavy interference from PU, which deteriorates its quality-of-service (QoS). Nevertheless, since the interference is an RF signal, it can be leveraged as a potential source of energy. For example, under the time switching policy, in the energy harvesting phase of a slot, the interference can be utilized as a useful energy source. This could subdue the harmful effect of the interference at the energy-constrained relay due to the additional energy that can be used to transmit with a higher power (provided it satisfies PU's interference threshold) to achieve a better QoS.

To this end, this paper analyzes the throughput performance of the secondary network while exploiting the primary interference as an energy source in addition to ST's signals. We use the outage probability and the ergodic capacity to characterize the QoS of SU in delay-limited and delay-tolerant transmission modes, respectively, whereas, the QoS of PU is characterized by its outage probability.

\subsection{Contributions}
This paper makes the following contributions.
\begin{itemize}
\item For both non-EH ST and EH ST cases, under the interference plus noise, interference dominant, and noise dominant scenarios, we derive exact analytical expressions for SU's outage probability and ergodic capacity, provided PU's outage probability remains below a threshold. We show that, due to the extra acquired energy, the interference-aided energy harvesting improves SU's average throughput performance compared to the case where the interference is treated as an unwanted signal in EH phase.
\item We take into account the impact of the energy harvesting activation threshold---characterized by the power outage probability---on SU's average throughput.  
\item To gain design insights, we study the effects of different system parameters such as the primary's transmit power, primary's outage threshold, number of primary transceivers, and peak power constraints on SU's throughput performance. We show that the peak power constraint is a key factor in deciding the optimal energy harvesting time.
\end{itemize}

\subsection{Related Work}
The benefits of RF energy harvesting in wireless networks were shown in~\cite{lu}. In the framework of cognitive radio with non-cooperative communications (i.e., without relays), authors in~\cite{rakovic_icc, zheng_el, zheng_sensor} presented the use of PU's RF signals to power CR networks. Specifically, in~\cite{rakovic_icc}, authors calculated the average achievable rate for a secondary direct link in the presence of a primary link where the ST used primary signals to harvest energy using a time switching policy. In a similar model, the reference~\cite{zheng_el} aimed to maximize the secondary achievable throughput and revealed an inherent energy harvesting-throughput trade-off. This work was further extended to the code-division multiple access (CDMA) framework in~\cite{zheng_sensor}, where multiple SUs that harvested energy from RF primary signals communicated with an access point.

In cognitive radio, using energy harvesting for energy-limited relays, authors in~\cite{sanket,van,yang_2016} showed that SUs could achieve substantial performance gains without battery recharging or replacement. In~\cite{sanket}, authors investigated a trade-off between the primary interference constraint and the energy constraint due to EH nature of relays and found the region of dominance for each of the constraints. The EH model considered in \cite{sanket} was a generic one and did not assume any specific source of energy. In \cite{van}, authors considered the RF EH model for secondary relays, where relays harvested energy from ST's information signals using a time switching policy. Under the instantaneous interference constraint imposed by the primary destination, authors in \cite{van} derived an outage probability expression for a secondary link. In \cite{yang_2016}, under the spectrum sharing with a single PU, authors obtained an analytical expression for the outage probability of the secondary network where an EH secondary relay used a power splitting policy to harvest energy. The works in~\cite{van,yang_2016} assumed no primary interference at secondary receivers and considered energy harvesting from only ST's signals. On the other hand, we analyze the performance of cognitive relays in the presence of the interference from multiple PUs and exploit such an interference as a useful energy source in addition to ST's signals. In~\cite{mousa}, authors analyzed the performance of the relay-assisted secondary communication in the presence of a single PU with an instantaneous interference constraint, where the PU interference was the sole source of energy for the secondary network. The reference~\cite{zheng_cl_16} studied the outage performance of multi-hop cognitive relaying where the relays harvested energy from a power beacon.

The work in~\cite{liu_2016} is similar in spirit to our work. Authors in~\cite{liu_2016} considered the interference from multiple PUs where both the ST and the SR harvested energy from primary signals. Their performance analysis focused on the interference dominant scenario where the noise is neglected when compared to the interference. Also in their framework, the primary destinations imposed an instantaneous interference constraint. In contrast, in our framework, the primary network imposes an outage constraint to guarantee a certain reliability to its users. In addition, we have considered interference plus noise scenario, interference dominant scenario, and noise dominant scenario. These scenarios can be used to characterize the proximity of primary transmitters to the secondary network and the effects of their transmit powers on the secondary network. For example, if primary transmitters are very close to the secondary network or their transmit powers are relatively large, the primary interference dominates the noise and we have an interference dominant scenario. On the other hand, if primary transmitters are located far from the secondary network or they transmit at relatively small powers, the noise dominates the primary interference. In addition to the framework considered in~\cite{liu_2016}, we also study another framework where the secondary relay harvests energy from the signals received from ST in addition to the primary interference. This reduces the dependence of the secondary network on the primary network for the energy, provides a more reliable source of energy as ST belongs to the same network, and emphasizes that the primary interference can be useful as it provides extra energy for the secondary transmission.\footnote{We refer readers to~\cite{he, zhong_2015} for the use of RF interference as an energy harvesting source in a non-cognitive radio setup.} We have compared these two frameworks and shown that harvesting energy from both the ST and the primary interference boosts the secondary throughput performance compared to harvesting energy either only from ST or only from primary interference. Another important feature that contrasts our paper from~\cite{liu_2016} is that we have depicted the effect of the peak power constraint on the energy harvesting time and the secondary throughput in detail, which provides additional insights into the system design (see Figs.~\ref{fig:opt_alpha_delay_limited}-\ref{fig:opt_alpha_EH_ST1}).

\subsection{Organization of the Paper}
\begin{itemize}
\item In Section~\ref{sec:sys_mod}, we describe the system model and the channel model for the non-EH ST case.
\item In Section~\ref{sec:max_pow}, we derive closed-form expressions for the maximum allowed transmit powers for ST and SR under the primary outage constraints and the peak power constraints.
\item In Section~\ref{sec:relay_prot}, we propose a time switching based energy harvesting policy at SR for interference plus noise and noise dominant scenarios.
\item In Section~\ref{sec:new}, we derive analytical expressions for the average SU throughput in interference plus noise scenario.
\item In Section~\ref{sec:derivations}, we derive analytical expressions for the average SU throughput in noise dominant scenario.
\item In Section~\ref{sec:EH_ST}, we extend our throughput analysis for the case where ST is also an EH node.
\item In Section~\ref{sec:results}, we discuss analytical and simulation results for both non-EH and EH ST cases.
\item In Section~\ref{sec:conc}, we provide conclusions and future directions.
\end{itemize}

\begin{figure}
\centering
\includegraphics[scale=0.25]{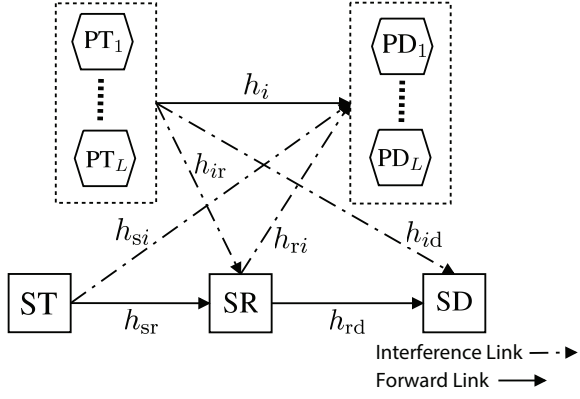}
\caption{Secondary communication via an interference-aided EH relay in spectrum sharing.}
\label{fig:syst}
\end{figure}

\section{System and Channel Models}
\label{sec:sys_mod}
As shown in Fig.~\ref{fig:syst}, we consider a primary network that consists of $L$ pairs of primary transmitters (PTs) and primary destinations (PDs). A spectrum of bandwidth $B$~$\mathrm{Hz}$ is divided equally among $L$ primary links, i.e., each primary link is allocated a bandwidth of $B/L$~$\mathrm{Hz}$. The secondary network consists of a secondary transmitter (ST) which communicates with a secondary destination (SD) via an energy harvesting decode-and-forward (DF) secondary relay (SR). All nodes have a single antenna. We focus on the underlay mode, where PUs are active during the operation of SUs and share the spectrum with the secondary network provided that the QoS of each primary link is maintained above a given threshold.

Let $h_{i}$, $h_{{\mathrm{sr}}}$, $h_{{\mathrm{rd}}}$, $h_{{\mathrm{s}}i}$, $h_{{\mathrm{r}}i}$, $h_{i\mathrm{r}}$, and $h_{i\mathrm{d}}$ denote the channel coefficients of $i$th primary link PT$_{i}$-PD$_{i}$ ($i = 1, 2, \dotsc, L$), ST-SR, SR-SD, ST-PD$_i$, SR-PD$_{i}$, PT$_i$-SR, and PT$_i$-SD links, respectively.
All channels are independent of each other and experience frequency-flat Rayleigh block fading. The instantaneous channel power gains are exponentially distributed random variables (RVs). Let us denote the mean channel power gain of $|h_{\mathrm{k}}|^{2}$ by $\lambda_{\mathrm{k}}$, where $\mathrm{k} \in \lbrace{i, \mathrm{sr}}, {\mathrm{rd}}, {\mathrm{s}}i, {\mathrm{r}}i, i\mathrm{r},  i\mathrm{d}\rbrace$. For simplicity, we consider the channels between PT and PD are identically distributed, i.e., $\lambda_{i}= \lambda_{{\mathrm{pp}}}$; interference channels from PTs to a node and vice-versa are also identically distributed, i.e., $\lambda_{i\mathrm{r}} = \lambda_{{\mathrm{p}}\mathrm{r}}$, $\lambda_{i\mathrm{d}} = \lambda_{{\mathrm{p}}\mathrm{d}}$, $\lambda_{{\mathrm{s}i}} = \lambda_{{\mathrm{s}}\mathrm{p}}$, and  $\lambda_{{\mathrm{r}i}} = \lambda_{{\mathrm{r}}\mathrm{p}}$. Due to limited feedback resources, we assume the availability of the knowledge of mean channel power gains for PT$_i$-PD$_i$, ST-PD$_i$, and SR-PD$_i$ links, as in~\cite{zou,peter:2013,rakovic_icc}. The mean channel power gains need only the statistical channel knowledge, which can be obtained by observing the channel for a sufficient time, and requires infrequent updates. On the other hand, SR and SD have the knowledge of instantaneous channels gains for the respective receiving links, i.e., for ST-SR and PT$_i$-SR links at SR and for SR-SD and PT$_i$-SD links at SD.\footnote{Since an $i$th PT transmits with a known constant power, the channel power gains on PT$_i$-SR and PT$_i$-SD links can be estimated using the received signals at SR and SD, respectively, as discussed in~\cite{rakovic_icc,zheng_sensor}.}

Assuming no direct link between ST and SD due to high attenuation~\cite{aissa_TWC, nasir,krik,yener1}, the secondary communication happens over two hops. In the first hop, ST transmits to SR, while in the second hop, SR forwards the received information to SD after decoding. The SR is an EH node which is capable of harvesting energy from the received RF signals. The SR may use some part of the received information signal to gather energy required to forward the information to SD. As the primary and secondary networks transmit simultaneously, SR experiences RF interference from $L$ PTs. Thus SR can harvest additional energy from the primary interference in the energy harvesting phase and convert it into a useful energy source. The ST and PTs are conventional nodes with constant power supply (e.g., battery). We shall consider the case where ST is also an EH node in Section~\ref{sec:EH_ST}.

\section{Maximum Allowed Secondary Transmit Powers}
\label{sec:max_pow}
The outage constraints at PDs govern the maximum transmit powers of ST and SR, i.e., ST and SR should limit their transmit powers so that the outage probability of each primary link remains below a given threshold. Let us denote the maximum allowed transmit powers of ST and SR due to primary outage constraints as $P_{\mathrm{ST}}$ and $P_{\mathrm{SR}}$, respectively. Then given the constant transmit power of PT ($P_{\mathrm{p}}$) and the interference from ST, the outage probability of $i$th primary link is
\begin{equation}
\mathrm{P}^{i}_{\mathrm{p, out, ST}} = \mathbb{P}\left(\frac{B}{L}\log_{2}\left(1+ \gamma_{\mathrm{PD}_i}\right)\leq R_\mathrm{p}\right) \leq \Theta_{\mathrm{p}},
\label{eq:P_out_st}
\end{equation}
where $\mathbb{P}(\cdot)$ denotes the probability, $\gamma_{\mathrm{PD}_i} = \frac{P_{\mathrm{p}}|h_{i}|^{2}}{P_{\mathrm{ST}}|h_{\mathrm{s}i}|^{2} + N_0}$ is the signal-to-interference noise ratio (SINR) at PD$_{i}$ with $N_0$ being the additive white Gaussian noise (AWGN) power at PDs, $R_\mathrm{p}$ is the desired primary rate for each primary link, and $\Theta_{\mathrm{p}}$ is the primary outage threshold for each PD. We assume that the codewords at transmitters (both primary and secondary) are drawn from a Gaussian codebook. Ensuring that the outage probability of the primary link having the worst SINR among $L$ links stays below $\Theta_{\mathrm{p}}$, we can express the primary outage constraint as
\begin{equation}
\mathrm{P}_{\mathrm{p, out, ST}} = \mathbb{P}\left(\min_{i = 1, 2, \dotsc, L} \mathrm{P}^{i}_{\mathrm{p, out, ST}}\right) \leq \Theta_{\mathrm{p}}.
\label{eq:Ppout}
\end{equation}
From \eqref{eq:P_out_st} and using the independence of $|h_{i}|^{2}$ and $|h_{\mathrm{s}i}|^{2}$, it follows that
\begin{align}
\mathrm{P}_{\mathrm{p, out, ST}}  = 1 - \prod_{i = 1}^{L} \left(1-\mathbb{P}\left( \frac{P_{\mathrm{p}}|h_{i}|^{2}}{P_{\mathrm{ST}}|h_{\mathrm{s}i}|^{2} + N_0} \leq \zeta_{\mathrm{p}}\right)\right),
\label{eq:Ppout2}
\end{align}
where $\zeta_{\mathrm{p}} = 2^{LR_\mathrm{p}/B} - 1$.
\begin{lemma}
Under primary outage constraints, ST's maximum transmit power is
\begin{equation}
P_{\mathrm{ST}} = \frac{P_{\mathrm{p}}\lambda_{\mathrm{pp}}}{\zeta_{\mathrm{p}} \lambda_{\mathrm{sp}}}\left(\frac{\mathcal{A}}{(1-\Theta_{\mathrm{p}})^{\frac{1}{L}}}- 1\right)^{+},
\label{eq:PST}
\end{equation}
where $\mathcal{A} = \exp\left(-\frac{\zeta_{\mathrm{p}}N_0B}{L P_{\mathrm{p}}\lambda_{\mathrm{pp}}}\right)$ and $(x)^{+} = \max(x, 0)$.
\end{lemma}
\begin{IEEEproof}
See Appendix \ref{sec:PST}.
\end{IEEEproof}
Similarly, we can write the maximum transmit power at SR as
\begin{equation}
P_{\mathrm{SR}} =\frac{P_{\mathrm{p}}\lambda_{\mathrm{pp}}}{\zeta_{\mathrm{p}} \lambda_{\mathrm{rp}}}\left(\frac{\mathcal{A}}{(1-\Theta_{\mathrm{p}})^{\frac{1}{L}}}- 1\right)^{+}.
\label{eq:Ppout7}
\end{equation}
Besides primary outage constraints, at both ST and SR, we impose peak power constraints $P_{\mathrm{t,ST}}$ and $P_{\mathrm{t,SR}}$, respectively. Then the maximum transmit powers at ST and SR respectively become
\begin{equation}
P_{\mathrm{Sm}} = \min\left(P_{\mathrm{ST}}, P_{\mathrm{t,ST}}\right),
\label{eq:ST}
\end{equation} 
and
\begin{equation}
P_{\mathrm{Rm}} = \min\left(P_{\mathrm{SR}}, P_{\mathrm{t,SR}}\right).
\label{eq:SR}
\end{equation}
Note that the maximum transmit power of SR given by \eqref{eq:SR} takes into account only the primary's outage constraints. In the next section, we shall calculate the maximum transmit power of SR that also takes into account the energy harvesting process. For non-EH ST case, the maximum transmit power of ST depends only on primary's outage constraints.

\begin{figure}
\centering
\includegraphics[scale=0.25]{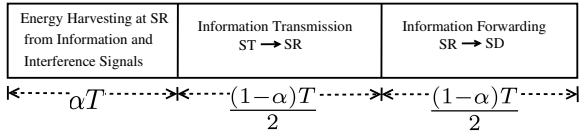}
\caption{Time switching protocol for the interference-aided energy harvesting and information processing at SR.}
\label{fig:protocol}
\end{figure}

\section{Relaying Protocol at SR}
\label{sec:relay_prot}

At SR, we adopt a time switching protocol to harvest energy from received RF signals as shown in Fig.~\ref{fig:protocol}. Under this protocol, at the start of a slot, for $\alpha T$ duration ($\alpha \in (0, 1)$), SR harvests energy from received RF signals, where $T$ is the duration of a slot of the secondary communication. The remaining time slot is divided into two sub-slots, each of duration $(1-\alpha)T/2$. In the first sub-slot, ST transmits information to SR; in the next sub-slot, SR forwards the information to SD. 
\subsection{Interference plus Noise Scenario}
Under this scenario, the received RF signals at relay include the signal from ST and the interference signals from $L$ PTs. The relay can boost its harvested energy by using the primary interference as an energy source. To activate the energy harvesting circuitry at the relay, the received power should be larger than the activation threshold $P_{\mathrm{th}}$. The threshold $P_{\rm{th}}$ depends on factors such as the type of the energy harvesting circuitry (linear or non-linear) and the frequency of incoming RF signals. In general, the threshold $P_{\mathrm{th}}$ ranges between $-30~\mathrm{dBm}$ to $-10~\mathrm{dBm}$~\cite{lu}. If the received power is smaller than $P_{\mathrm{th}}$, it leads to a power outage event, which in turn causes a secondary outage event. We call the probability of the occurrence of a power outage event as the \textit{power outage probability} and give its expression in the following proposition.
\begin{proposition}
\label{prop:power_outage_start}
Given the energy harvesting circuitry activation threshold $P_{\mathrm{th}}$, we write the power outage probability as
\begin{align}
P_{\mathrm{e,out}} \!= \!\left\{
  \begin{array}{l l}\!\!\!
  \frac{\gamma\left(L, \frac{P_{\mathrm{th}}}{P_{\mathrm{p}}\lambda_{\mathrm{pr}}}\right)}{\Gamma(L)} - \frac{e^{-\frac{P_{\mathrm{th}}}{P_{\mathrm{Sm}}\lambda_{\mathrm{sr}}}}\gamma(L,\omega P_{\mathrm{th}})}{\Gamma(L)(\omega P_{\mathrm{p}} \lambda_{\mathrm{pr}})^{L}}, &\quad  \hspace*{-5mm}P_\mathrm{p}\lambda_{\mathrm{pr}} \neq P_{\mathrm{Sm}}\lambda_{\mathrm{sr}},\\
    \!\!\!\frac{\gamma\left(L+1, \frac{P_{\mathrm{th}}}{P_{\mathrm{p}}\lambda_{\mathrm{pr}}} \right)}{\Gamma(L+1)}, & \quad \hspace*{-5mm} P_\mathrm{p}\lambda_{\mathrm{pr}} = P_{\mathrm{Sm}}\lambda_{\mathrm{sr}},\\
  \end{array} \right.
\label{eq:peout_basic}
\end{align}
where $\omega$ is
\begin{equation}
\omega = \frac{1}{P_{\mathrm{p}}\lambda_{\mathrm{pr}}} - \frac{1}{P_{\mathrm{Sm}}\lambda_{\mathrm{sr}}},
\label{eq:omegag}
\end{equation}
$\gamma(a,b) = \int_{0}^{b} t^{a-1}\exp(-t)\mathrm{d}t$ is the lower incomplete gamma function, and $\Gamma(a) = \int_{0}^{\infty} t^{a-1}\exp(-t)\mathrm{d}t$ is the gamma function.
\end{proposition}
\begin{IEEEproof}
See Appendix~\ref{app:power_outage_start}.
\end{IEEEproof}
Given that the energy harvesting circuitry at SR is active, when ST transmits with power $P_{\mathrm{Sm}}$ and each PT transmits with power $P_{\mathrm{p}}$, the energy harvested at SR in $\alpha T$ duration is given by
\begin{equation}
E_{\mathrm{SR,H}} = (\alpha T)\delta \left(P_{\mathrm{Sm}}|h_{\mathrm{sr}}|^{2} + \sum_{i = 1}^{L} P_{\mathrm{p}}|h_{i\mathrm{r}}|^2  \right),
\label{eq:harv_energy}
\end{equation}
where $\delta$, with $0 \leq \delta \leq 1$, is the energy conversion efficiency factor. The SR uses harvested energy to forward the information to SD. Then given the amount of harvested energy, the transmit power of SR can be given by\footnote{Usually, the energy consumption by the circuitry of SR in information processing is negligible compared to that in transmission~\cite{nasir,nasir2}. Hence we assume that SR uses all the harvested energy for the transmission purpose.}
\begin{equation}
P_{\mathrm{SR, H}} = \frac{2 E_{\mathrm{SR,H}}}{(1-\alpha)T} = \frac{2\delta \alpha}{1-\alpha}\left(P_{\mathrm{Sm}}|h_{\mathrm{sr}}|^{2} +  \sum_{i = 1}^{L}P_{\mathrm{p}}|h_{i\mathrm{r}}|^2 \right).
\label{eq:harv_power}
\end{equation}
By incorporating primary outage constraints and the peak power constraint at SR, the maximum transmit power of SR is
\begin{equation}
P_{\mathrm{R}} = \min\left(P_{\mathrm{SR, H}}, P_{\mathrm{Rm}}\right),
\label{eq:SR_fin}
\end{equation}
where $P_{\mathrm{Rm}}$ is given by \eqref{eq:SR}. We can write the SINR at relay as
\begin{equation}
\gamma_{\mathrm{SR}} = \frac{P_{\mathrm{Sm}} |h_{\mathrm{sr}}|^{2}}{\displaystyle \sum_{i = 1}^{L}P_{\mathrm{p}}|h_{i \mathrm{r}}|^2 + N_{0,\mathrm{r}}},
\label{eq:SINR_SR}
\end{equation}
where $N_{0,\mathrm{r}}$ is the noise power at relay. Similarly, the SINR at the secondary destination is given by
\begin{align}
\gamma_{\mathrm{SD}} &= \frac{P_{\mathrm{R}} |h_{\mathrm{rd}}|^{2}}{\displaystyle \sum_{i = 1}^{L}P_{\mathrm{p}}|h_{i \mathrm{d}}|^2 + N_{0,\mathrm{d}}} \nonumber \\
&=\frac{\min\left( \frac{2\delta \alpha}{1-\alpha}\left(P_{\mathrm{Sm}}|h_{\mathrm{sr}}|^{2} +  \sum_{i = 1}^{L}P_{\mathrm{p}}|h_{i\mathrm{r}}|^2 \right), P_{\mathrm{Rm}}\right) |h_{\mathrm{rd}}|^{2}}{\displaystyle \sum_{i = 1}^{L}P_{\mathrm{p}}|h_{i \mathrm{d}}|^2 + N_{0,\mathrm{d}}},
\label{eq:SINR_SD}
\end{align}
where $N_{0,\mathrm{d}}$ is the noise power at the secondary destination. Note that $\gamma_{\mathrm{SR}}$ given by \eqref{eq:SINR_SR} and $\gamma_{\mathrm{SD}}$ given by \eqref{eq:SINR_SD} are dependent RVs due to the presence of the common RV $\sum_{i = 1}^{L}P_{\mathrm{p}}|h_{i \mathrm{r}}|^2$.  The end-to-end SINR for DF relaying is given as
\begin{equation}
\gamma_{\mathrm{e2e}} = \min(\gamma_{\mathrm{SR}}, \gamma_{\mathrm{SD}}).
\label{eq:e2e_SINR}
\end{equation}

\subsection{Interference Dominant Scenario}

In the case of interference dominant scenario where the interference power due to primary transmissions is much larger than the noise power, one can obtain signal-to-interference ratio (SIR) by neglecting the noise powers in \eqref{eq:SINR_SR} and \eqref{eq:SINR_SD}. The harvested energy and the harvested power at relay remain the same as that of the interference plus noise scenario.

\subsection{Noise Dominant Scenario}
In this scenario, the noise power dominates the primary interference power. This case occurs when primary transmitters are located far from the secondary network or transmit at small powers. Hence the energy harvested by SR mainly comes from ST's signals. In the following proposition, we give an expression of the power outage probability.
\begin{proposition}
Given the energy harvesting circuitry activation threshold $P_{\mathrm{th}}$, we write the power outage probability as
\begin{align}
P_{\mathrm{e,out}} = 1 - \exp\left( -\frac{P_{\mathrm{th}}}{P_{\mathrm{Sm}}\lambda_{\mathrm{sr}}}\right).
\label{eq:peout_N_basic}
\end{align}
\end{proposition}
\begin{IEEEproof}
Since the primary interference power is negligible, the received power at relay can be given as
\begin{align}
P_{\mathrm{rec}} = P_{\mathrm{Sm}} |h_{\mathrm{sr}}|^{2}.
\end{align}
The power outage probability follows as
\begin{align}
P_{\mathrm{e,out}} &= \mathbb{P}(P_{\mathrm{rec}} < P_{\mathrm{th}}) \nonumber \\
&= \mathbb{P}\left(|h_{\mathrm{sr}}|^{2} < \frac{P_{\mathrm{th}}}{P_{\mathrm{Sm}}}\right)  \nonumber \\
&= 1 - \exp\left( -\frac{P_{\mathrm{th}}}{P_{\mathrm{Sm}}\lambda_{\mathrm{sr}}}\right).
\end{align}
\end{IEEEproof}
Given that the energy harvesting circuitry at SR is active, the energy harvested by SR is
\begin{equation}
E_{\mathrm{SR,H}} = (\alpha T)\delta P_{\mathrm{Sm}}|h_{\mathrm{sr}}|^{2}.
\label{eq:harv_energy_noise_only}
\end{equation}
Accordingly, the transmit power at SR is
\begin{equation}
P_{\mathrm{SR, H}} = \frac{2 E_{\mathrm{SR,H}}}{(1-\alpha)T} = \frac{2\delta \alpha}{1-\alpha}P_{\mathrm{Sm}}|h_{\mathrm{sr}}|^{2}.
\label{eq:harv_power_noise_only}
\end{equation}
Including primary outage constraints and the peak power constraint at SR, the maximum transmit power $P_{\mathrm{R}}$ at SR can be given by \eqref{eq:SR_fin}. We can then write the signal-to-noise ratio (SNR) at relay as
\begin{equation}
\gamma_{\mathrm{SR}} = \frac{P_{\mathrm{Sm}} |h_{\mathrm{sr}}|^{2}}{ N_{0,\mathrm{r}}},
\label{eq:SNR_SR_noise}
\end{equation}
Similarly the SNR at the secondary destination is given by
\begin{align}
\gamma_{\mathrm{SD}} &= \frac{P_{\mathrm{R}} |h_{\mathrm{rd}}|^{2}}{N_{0,\mathrm{d}}}  = \frac{{\min\left(\frac{2\delta \alpha}{1-\alpha}P_{\mathrm{Sm}}|h_{\mathrm{sr}}|^{2}, P_{\mathrm{Rm}}\right)|h_{\mathrm{rd}}|^{2}}}{N_{0,\mathrm{d}}}.
\label{eq:SNR_SD_noise}
\end{align}
Note that $\gamma_{\mathrm{SR}}$ (given by \eqref{eq:SNR_SR_noise}) and $\gamma_{\mathrm{SD}}$ (given by \eqref{eq:SNR_SD_noise}) are dependent RVs due to the presence of the common RV $P_{\mathrm{Sm}} |h_{\mathrm{sr}}|^{2}$. Hereafter, without loss of generality, we assume that the duration of a time-slot is $T = 1$ and $N_{0,\mathrm{r}} = N_{0,\mathrm{d}} = N_0$.

\section{Throughput Analysis: Interference plus Noise Scenario}
\label{sec:new}
In this section, we focus on the interference plus noise scenario and derive analytical expressions for the average throughput for both the delay-limited and the delay-tolerant secondary transmissions.

\subsection{Delay-Limited Secondary Transmission}
In the delay-limited transmission mode, the outage probability characterizes SU's performance. Under this mode, ST transmits with a fixed rate $R_{\mathrm{s}}$ bits/s/Hz. Then the SU throughput is obtained by calculating the outage probability $P_{\mathrm{s,out}}$ for the rate $R_{\mathrm{s}}$. Let us denote $\zeta_{\mathrm{s}}$ as the threshold SINR required to detect the received information correctly at SD. Then we have $\zeta_{\mathrm{s}} = 2^{\frac{R_{\mathrm{s}}}{B}} -1 $. In this case, the average achieved throughput can be given as
\begin{equation} 
\mathcal{R}_{\mathrm{s}} = \frac{1-\alpha}{2}(1-P_{\mathrm{e,out}})(1-P_{\mathrm{s,out}})R_{\mathrm{s}},
\label{eq:thro_IN}
\end{equation}
where the term $1-P_{\mathrm{e,out}}$ denotes the probability that the energy harvesting circuitry at relay is active. The secondary outage probability $P_{\mathrm{s,out}}$ is the probability that the end-to-end SINR $\gamma_{\mathrm{e2e}}$ given by \eqref{eq:e2e_SINR} is below the threshold SINR $\zeta_{\mathrm{s}}$, i.e.,
\begin{align}
P_{\mathrm{s,out}}(\zeta_{\mathrm{s}}) &= \mathbb{P}\left(\gamma_{\mathrm{e2e}} < \zeta_{\mathrm{s}}\right)  = \mathbb{P}\left(\min(\gamma_{\mathrm{SR}}, \gamma_{\mathrm{SD}}) < \zeta_{\mathrm{s}}\right) \nonumber \\
&= 1 - \mathbb{P}\left(\gamma_{\mathrm{SR}} \geq \zeta_{\mathrm{s}}, \gamma_{\mathrm{SD}} \geq \zeta_{\mathrm{s}}\right),
\label{eq:Psout_IN_basic}
\end{align}
where $\gamma_{\mathrm{SR}}$ and $\gamma_{\mathrm{SD}}$ are SINRs at SR and SD, respectively, and are given by \eqref{eq:SINR_SR} and \eqref{eq:SINR_SD}. The following proposition provides an analytical expression for $P_{\mathrm{s,out}}$.

\begin{proposition}
\label{prop:psout_IN}
For ST's fixed transmission rate $R_{\mathrm{s}}$, the secondary outage probability $P_{\mathrm{s,out}}$ is
\begin{equation}
P_{\mathrm{s,out}}(\zeta_{\mathrm{s}}) =  1 - \frac{1}{(P_{\mathrm{p}}\lambda_{\mathrm{pd}})^{L}}(I_1 + I_2 + I_3),
\label{eq:Psout_IN}
\end{equation}
where
\begin{subequations}
\begin{align}
I_1 &= \int_{u = 0}^{\delta_1} \int_{x = \frac{\zeta_{\mathrm{s}}(u+N_0)}{P_{\mathrm{Sm}}}}^{\frac{P_{\mathrm{Rm}} -\theta u}{\theta P_{\mathrm{Sm}}}} f_{X}(x)f_{U}(u)\exp\left(-\frac{\zeta_{\mathrm{s}}N_0}{\theta (P_{\mathrm{Sm}}x + u)\lambda_{\mathrm{rd}}}\right) \nonumber \\
& \times \left(\frac{1}{P_{\mathrm{p}}\lambda_{\mathrm{pd}}} + \frac{\zeta_{\mathrm{s}}}{\theta(P_{\mathrm{Sm}}x +u)\lambda_{\mathrm{rd}}}\right)^{-L}\mathrm{d}x \,\mathrm{d}u, 
\end{align}
\begin{align}
I_2 &= \frac{\gamma(L,\omega \delta_1) \exp\left(-\frac{\zeta_{\mathrm{s}}N_0}{P_{\mathrm{Rm}}\lambda_{\mathrm{rd}}}-\frac{P_{\mathrm{Rm}}}{\theta P_{\mathrm{Sm}}\lambda_{\mathrm{sr}}}\right)}{\Gamma(L)(P_{\mathrm{p}}\lambda_{\mathrm{pr}})^{L} \omega^{L}\left(\frac{1}{P_{\mathrm{p}}\lambda_{\mathrm{pd}}} + \frac{\zeta_{\mathrm{s}}}{P_{\mathrm{Rm}}\lambda_{\mathrm{rd}}}\right)^{L}},
\label{eq:I22} 
\end{align}
\begin{align}
I_3 &= \frac{\Gamma(L,\varphi \delta_1) \exp\left(-\frac{\zeta_{\mathrm{s}}N_0}{P_{\mathrm{Rm}}\lambda_{\mathrm{rd}}}-\frac{\zeta_{\mathrm{s}}N_0}{P_{\mathrm{Sm}}\lambda_{\mathrm{sr}}}\right)}{\Gamma(L)(P_{\mathrm{p}}\lambda_{\mathrm{pr}})^{L} \varphi^{L}\left(\frac{1}{P_{\mathrm{p}}\lambda_{\mathrm{pd}}} + \frac{\zeta_{\mathrm{s}}}{P_{\mathrm{Rm}}\lambda_{\mathrm{rd}}}\right)^{L}},
\label{eq:I33} 
\end{align}
with
\begin{align}
f_X(x) =\frac{1}{P_{\mathrm{Sm}}\lambda_{\mathrm{sr}}}\exp\left(-\frac{x}{P_{\mathrm{Sm}}\lambda_{\mathrm{sr}}}\right), \quad x \geq 0,
\end{align}
\begin{align}
f_U(u) = \frac{u^{L-1}\exp\left(-\frac{u}{P_{\mathrm{p}}\lambda_{\mathrm{pr}}}\right)}{\Gamma(L)(P_{\mathrm{p}}\lambda_{\mathrm{pr}})^{L}}, \quad u \geq 0,
\label{eq:fuu}
\end{align}
\begin{equation}
\delta_1 = \frac{P_{\mathrm{Rm}}-\zeta_{\mathrm{s}}\theta N_0}{(1+\zeta_{\mathrm{s}})\theta},
\end{equation}
\begin{equation}
\theta = \frac{2 \delta \alpha}{1-\alpha},
\end{equation}
\begin{equation}
\varphi = \frac{1}{P_{\mathrm{p}\lambda_{\mathrm{pr}}}} + \frac{\zeta_{\mathrm{s}}}{P_{\mathrm{Sm}}\lambda_{\mathrm{sr}}},
\end{equation}
$\Gamma(a,b) = \int_{b}^{\infty} t^{a-1}\exp(-t)\mathrm{d}t$ is the upper incomplete gamma function, and $\omega$ is given by \eqref{eq:omegag}.
\end{subequations}
\end{proposition}
\begin{IEEEproof}
See Appendix  \ref{sec:psout_IN}.
\end{IEEEproof}
Substituting $P_{\mathrm{s,out}}$ from \eqref{eq:Psout_IN} in \eqref{eq:thro_IN}, we obtain the average SU throughput.

\subsection{Delay-Tolerant Secondary Transmission}
\label{sec:del_tol_in}
Under this transmission mode, the ergodic capacity characterizes SU's performance. In particular, ST transmits with the rate that is less than the ergodic capacity. Under the delay-tolerant transmission mode, the codeword length is very large compared to the length of a block. Hence the code sees all the possible channel realizations, which makes the ergodic capacity an appropriate measure of SU's performance. In this case, we can express the average SU throughput as
\begin{equation}
\mathcal{R}_{\mathrm{s}} = \frac{1-\alpha}{2}(1-P_{\mathrm{e,out}})C_0,
\label{eq:thro_erg_IN}
\end{equation}
where $C_0$ denotes the ergodic capacity which is given as
\begin{align}
C_0 &= \mathbb{E}\left[\log_2(1+\gamma_{\mathrm{e2e}}) \right] = \int_{t=0}^{\infty} \log_2(1+\gamma_{\mathrm{e2e}})f_{\gamma_{\mathrm{e2e}}}(t)\mathrm{d}t,
\label{eq:erg_IN}
\end{align}
where $\mathbb{E}(\cdot)$ is the expectation operator, $\gamma_{\mathrm{e2e}}$ is a random variable given by \eqref{eq:e2e_SINR}, and $f_{\gamma_{\mathrm{e2e}}}(t)$ is the probability density function of $\gamma_{\mathrm{e2e}}$. Using the integration by parts, it follows that
\begin{align}
C_0 = \frac{1}{\ln 2}\int_{t=0}^{\infty} \frac{1}{1+t}(1-F_{\gamma_{\mathrm{e2e}}}(t))\mathrm{d}t,
\label{eq:erg_IN1}
\end{align}
where $F_{\gamma_{\mathrm{e2e}}}(t) = \mathbb{P}(\gamma_{\mathrm{e2e}} < t)$ denotes the cumulative distribution function (CDF) of $\gamma_{\mathrm{e2e}}$. Thus, using \eqref{eq:Psout_IN_basic}, $F_{\gamma_{\mathrm{e2e}}}(t)$ can be obtained from \eqref{eq:Psout_IN}, i.e., $F_{\gamma_{\mathrm{e2e}}}(t) = P_{\mathrm{s,out}}(t)$. Now substituting \eqref{eq:erg_IN1} in \eqref{eq:thro_erg_IN}, we obtain the average SU throughput.

\begin{remark}
For the interference dominant case, the power outage probability is the same as that of the interference and noise scenario (given by \eqref{eq:peout_basic}), while the secondary outage probability and the average throughput in both delay-limited and delay-tolerant modes can be obtained by setting $N_0 = 0$ in corresponding expressions for the interference and noise scenario.
\end{remark}

\section{Throughput Analysis: Noise Dominant Scenario}
\label{sec:derivations}
In this section, we focus on the noise dominant scenario where we derive analytical expressions for the average SU throughput for both the delay-limited and the delay-tolerant secondary transmissions. 

\subsection{Delay-Limited Secondary Transmission}
The following proposition gives an analytical expression of the secondary outage probability.
\begin{proposition}
\label{prop:psout_N}
For ST's fixed transmission rate $R_{\mathrm{s}}$, the secondary outage probability $P_{\mathrm{s,out}}$ is
\begin{equation}
\label{eq:binary11}
P_{\mathrm{s,out}}(\zeta_{\mathrm{s}}) = \left\{
  \begin{array}{l l}
    I_4, & \quad \text{if}~\frac{P_{\mathrm{Rm}}}{\theta P_{\mathrm{Sm}}} \geq \frac{N_0 \zeta_{\mathrm{s}}}{P_{\mathrm{Sm}}} \\
    I_5, & \quad \text{if}~\frac{P_{\mathrm{Rm}}}{\theta P_{\mathrm{Sm}}} < \frac{N_0 \zeta_{\mathrm{s}}}{P_{\mathrm{Sm}}},\\
  \end{array} \right.
\end{equation}
where 
\begin{align}
I_4 &= 1 - \lambda_{\mathrm{sr}}\Bigg[\frac{P_{\mathrm{Rm}}}{\theta P_{\mathrm{Sm}}}K_1\left(\frac{N_0 \zeta_{\mathrm{s}}}{\lambda_{\mathrm{rd}}P_{\mathrm{Rm}}}, \frac{P_{\mathrm{Rm}}}{\theta P_{\mathrm{Sm}}\lambda_{\mathrm{sr}}}\right) \nonumber \\
& \!- \frac{N_0 \zeta_{\mathrm{s}}}{P_{\mathrm{Sm}}}K_1\!\left(\!\frac{1}{\theta \lambda_{\mathrm{rd}}},\frac{N_0 \zeta_{\mathrm{s}}}{P_{\mathrm{Sm}}\lambda_{\mathrm{sr}}}\!\right)\!\!\Bigg] + \exp \!\left(\!-\frac{P_{\mathrm{Rm}}}{\theta P_{\mathrm{Sm}}\lambda_{\mathrm{sr}}} - \frac{N_0 \zeta_{\mathrm{s}}}{\lambda_{\mathrm{rd}}P_{\mathrm{Rm}}}\!\right),
\label{eq:I4N}
\end{align}
\begin{align}
I_5 &= 1 - \exp\left(-\left(\frac{N_0 \zeta_{\mathrm{s}}}{P_{\mathrm{Sm}}\lambda_{\mathrm{sr}}} + \frac{N_0 \zeta_{\mathrm{s}}}{P_{\mathrm{Rm}}\lambda_{\mathrm{rd}}}\right)\right),
\label{eq:I5N}
\end{align}
with 
\begin{equation}
\theta = \frac{2 \delta \alpha}{1-\alpha},
\end{equation}
and $K_\nu(x,y)$ is the incomplete Bessel function defined as
\begin{equation}
K_\nu(x,y) = \int_{1}^{\infty} \frac{\exp\left(-xt-\frac{y}{t}\right)}{t^{\nu+1}}\mathrm{d}t.
\end{equation}

\end{proposition}
\begin{IEEEproof}
See Appendix \ref{app:noise}.
\end{IEEEproof}

\subsection{Delay-Tolerant Secondary Transmission}
Similar to the interference and noise scenario (see Section \ref{sec:del_tol_in}), we can express the average SU throughput as
 \begin{equation}
\mathcal{R}_{\mathrm{s}} = \frac{1-\alpha}{2}(1-P_{\mathrm{e,out}})C_0,
\label{eq:thro_erg_N}
\end{equation}
where $P_{\mathrm{e,out}}$ is given by \eqref{eq:peout_N_basic} and $C_0$ is given by \eqref{eq:erg_IN1} with $F_{\gamma_{\mathrm{e2e}}}(t) = P_{\mathrm{s,out}}(t)$. Here $P_{\mathrm{s,out}}$ is given by \eqref{eq:binary11}.

\section{Energy Harvesting ST}
\label{sec:EH_ST}
In this section, we consider a scenario where ST along with SR is also an EH node. In this case, there is only one energy source, namely, the primary interference. For this scenario, a time slot is divided in three sub-slots. In the first sub-slot of duration $\alpha T$, both ST and SR harvest energy from primary's signals. Each of the next two sub-slots is of $\frac{1-\alpha}{2}T$ duration, where ST transmits information to SR in the second sub-slot, which SR forwards to SD in the third sub-slot.

\subsection{Power Outage Probability}
To activate the energy harvesting circuitry at ST and SR, the received power should be larger than the activation threshold $P_{\mathrm{th}}$. In this case, the power outage probability at ST is given by the following proposition.
\begin{proposition}
Given the energy activation threshold $P_{\mathrm{th}}$, the power outage probability at ST is
\begin{equation}
P_{\mathrm{e,out,ST}} = \frac{\gamma\left(L, \frac{P_{\mathrm{th}}}{P_{\mathrm{p}}\lambda_{\mathrm{ps}}}\right)}{\Gamma(L)}.
\label{eq:ppout_ST}
\end{equation}
\end{proposition}
\begin{IEEEproof}
Since ST harvests energy from primary's signals, the received power at ST is
\begin{equation}
P_{\mathrm{rec,ST}} = \sum_{i= 1}^{L} P_{\mathrm{p}}|h_{i{\mathrm{s}}}|^{2},
\end{equation}
where $|h_{i\mathrm{s}}|^{2}$ is Rayleigh channel power gain from $i$th PT to ST. The power outage probability follows as
\begin{align}
P_{\mathrm{e,out,ST}} &= \mathbb{P}(P_{\mathrm{rec,ST}} < P_{\mathrm{th}}) = \mathbb{P}\left(\sum_{i= 1}^{L} P_{\mathrm{p}}|h_{i\mathrm{s}}|^{2} < P_{\mathrm{th}}\right),
\label{eq:ppout_ST1}
\end{align} 
where $|h_{i\mathrm{s}}|^{2}$ is exponentially distributed. Here $\sum_{i= 1}^{L} P_{\mathrm{p}}|h_{i\mathrm{s}}|^{2}$ is a gamma RV with the shape parameter $L$ and the scale parameter $P_{\mathrm{p}}\lambda_{\mathrm{ps}}$ where $\lambda_{\mathrm{ps}}$ is the mean channel gain on the link between $i$th PT and ST. Hence we can write the power outage probability in \eqref{eq:ppout_ST1} as the CDF of the gamma RV, which is the required expression in \eqref{eq:ppout_ST}.
\end{IEEEproof}
Similarly, the power outage probability at SR is
\begin{align}
P_{\mathrm{e,out,SR}} &= \mathbb{P}(P_{\mathrm{rec,SR}} < P_{\mathrm{th}}) \nonumber \\
&= \mathbb{P}\left(\sum_{i= 1}^{L} P_{\mathrm{p}}|h_{i\mathrm{r}}|^{2} < P_{\mathrm{th}}\right) \nonumber \\
&=  \frac{\gamma\left(L, \frac{P_{\mathrm{th}}}{P_{\mathrm{p}}\lambda_{\mathrm{pr}}}\right)}{\Gamma(L)},
\label{eq:ppout_SR}
\end{align} 
where $P_{\mathrm{rec,SR}}$ is the power received at SR. Including the power outage probabilities at both ST and SR, the overall power outage probability $P_{\mathrm{e,out}}$ is given as 
\begin{equation}
P_{\mathrm{e,out}} = 1 -(1-P_{\mathrm{e,out,ST}})(1-P_{\mathrm{e,out,SR}}).
\end{equation}

\subsection{Energy Harvesting at ST and SR}

Given that the energy harvesting circuitry at ST is active, in the first sub-slot of the energy harvesting protocol, the energy harvested at ST is
\begin{align}
E_{\mathrm{ST,H}} = (\alpha T)\delta \sum_{i= 1}^{L} P_{\mathrm{p}}|h_{i\mathrm{s}}|^{2}.
\end{align} 
The ST uses this harvested energy to transmit information to SR for the duration of $\frac{1-\alpha}{2}$ with power
\begin{align}
P_{\mathrm{ST,H}} = \frac{2\alpha \delta}{1-\alpha} \sum_{i= 1}^{L} P_{\mathrm{p}}|h_{i\mathrm{s}}|^{2},
\label{eq:Pst_EH}
\end{align}
where we have assumed $T = 1$ without loss of generality.
At the same time, SR harvests energy from primary's interference signals, which is given by
\begin{align}
E_{\mathrm{SR,H}} = \alpha\delta \sum_{i= 1}^{L} P_{\mathrm{p}}|h_{i\mathrm{r}}|^{2}.
\end{align}
Then the transmit power of the relay to forward the information follows as
\begin{align}
P_{\mathrm{SR,H}} = \frac{2\alpha \delta}{1-\alpha} \sum_{i= 1}^{L} P_{\mathrm{p}}|h_{i\mathrm{r}}|^{2}.
\label{eq:Psr_EH}
\end{align}
\subsection{Delay-Limited Secondary Transmission}
At ST, considering the maximum allowed power $P_{\mathrm{ST}}$ due to primary outage constraints and the peak power constraint $P_{\mathrm{t,ST}}$ along with the harvested power, we can write its transmit power as
\begin{align}
P_{\mathrm{S}} = \min(P_{\mathrm{ST,H}},P_{\mathrm{Sm}}),
\end{align}  
where $P_{\mathrm{ST,H}}$ is given by \eqref{eq:Pst_EH} and $P_{\mathrm{Sm}} = \min(P_{\mathrm{ST}},P_{\mathrm{t,ST}})$ as given by \eqref{eq:ST}.

Similarly, at SR, considering the maximum allowed power $P_{\mathrm{SR}}$ due to primary outage constraints and the peak power constraint $P_{\mathrm{t,SR}}$ along with the harvested power, we can write its transmit power as
\begin{align}
P_{\mathrm{R}} = \min(P_{\mathrm{SR,H}},P_{\mathrm{Rm}}),
\end{align}  
where $P_{\mathrm{SR,H}}$ is given by \eqref{eq:Psr_EH} and $P_{\mathrm{Rm}} = \min(P_{\mathrm{SR}},P_{\mathrm{t,SR}})$ is given by \eqref{eq:SR}.

For the interference plus noise scenario, we can write SINR at the relay as
\begin{align}
\gamma_{\mathrm{SR}} &= \frac{P_{\mathrm{S}} |h_{\mathrm{sr}}|^{2}}{\displaystyle \sum_{i = 1}^{L}P_{\mathrm{p}}|h_{i \mathrm{r}}|^2 + N_{0,\mathrm{r}}} \nonumber \\
&=\frac{\min\left( \frac{2\delta \alpha}{1-\alpha}\left( \sum_{i = 1}^{L}P_{\mathrm{p}}|h_{i\mathrm{s}}|^2 \right), P_{\mathrm{Sm}}\right) |h_{\mathrm{sr}}|^{2}}{\displaystyle \sum_{i = 1}^{L}P_{\mathrm{p}}|h_{i \mathrm{r}}|^2 + N_{0,\mathrm{r}}}.
\label{eq:SINR_SR_st_eh}
\end{align}
Similarly, SINR at the secondary destination is given by
\begin{align}
\gamma_{\mathrm{SD}} =\frac{\min\left( \frac{2\delta \alpha}{1-\alpha}\left( \sum_{i = 1}^{L}P_{\mathrm{p}}|h_{i\mathrm{r}}|^2 \right), P_{\mathrm{Rm}}\right) |h_{\mathrm{rd}}|^{2}}{\displaystyle \sum_{i = 1}^{L}P_{\mathrm{p}}|h_{i \mathrm{d}}|^2 + N_{0,\mathrm{d}}}.
\label{eq:SINR_SD_st_eh}
\end{align}
The following proposition provides an analytical expression for the secondary outage probability. 
\begin{proposition}
\label{prop:ST_EH_out}
For ST's fixed transmission rate $R_{\mathrm{s}}$, the secondary outage probability $P_{\mathrm{s,out}}$ is
\begin{align}
P_{\mathrm{s,out}}(\zeta_{\mathrm{s}})  = 1- \int_{u = 0}^{\infty} \left[(T_1 + T_2)T_3\right] f_{U}(u)~\mathrm{d}u,
\label{eq:ST_EH_out}
\end{align}
where 
\begin{equation}
T_1 = \frac{(P_{\mathrm{Sm}})^{L}}{\Gamma(L)(\theta P_{\mathrm{p}}\lambda_{\mathrm{ps}})^{L}}K_{L}\left(\frac{\theta (u+N_{\mathrm{0,r}})\zeta_{\mathrm{s}}}{\theta P_{\mathrm{Sm}}\lambda_{\mathrm{sr}}},\frac{P_{\mathrm{Sm}}}{\theta P_{\mathrm{p}} \lambda_{\mathrm{ps}}}\right),
\label{eq:T1_fin}
\end{equation}
\begin{equation}
T_2 = \frac{\exp\left(-\frac{(u + N_{0,\mathrm{r}})\zeta_{\mathrm{s}}}{P_{\mathrm{Sm}}\lambda_{\mathrm{sr}}}\right)\Gamma\left(L,\frac{P_{\mathrm{Sm}}}{\theta P_{\mathrm{p}} \lambda_{\mathrm{ps}}}\right)}{\Gamma(L)},
\end{equation}
\begin{equation}
T_3 = \frac{\exp\left(-\frac{(\zeta_{\mathrm{s}}N_{0,\mathrm{d}})}{\min(\theta u, P_{\mathrm{Rm}})\lambda_{\mathrm{rd}}}\right)\left(\frac{1}{P_{\mathrm{p}}\lambda_{\mathrm{pd}}} + \frac{\zeta_{\mathrm{s}}}{\min(\theta u, P_{\mathrm{Rm}})\lambda_{\mathrm{rd}}}\right)^{-L}}{(P_{\mathrm{p}}\lambda_{\mathrm{pd}})^{L}},
\end{equation}
and $f_{U}(u)$ is given by \eqref{eq:fuu}.

\end{proposition}
\begin{IEEEproof}
See Appendix~\ref{app:ST_EH}.
\end{IEEEproof}
The average SU throughput is
\begin{align}
\mathcal{R}_{\mathrm{s}} = \frac{1-\alpha}{2}(1-P_{\mathrm{e,out,ST}})(1-P_{\mathrm{e,out,SR}})(1-P_{\mathrm{s,out}})R_{\mathrm{s}},
\end{align}
where $P_{\mathrm{e,out,ST}}$ and $P_{\mathrm{e,out,SR}}$ are given by \eqref{eq:ppout_ST} and \eqref{eq:ppout_SR}, respectively.

\subsection{Delay-Tolerant Secondary Transmission}
In this case, the average SU throughput is
 \begin{equation}
\mathcal{R}_{\mathrm{s}} = \frac{1-\alpha}{2}(1-P_{\mathrm{e,out,ST}})(1-P_{\mathrm{e,out,SR}})C_0,
\label{eq:thro_erg_N1}
\end{equation}
where $P_{\mathrm{e,out}}$ is given by \eqref{eq:peout_N_basic} and $C_0$ is given by \eqref{eq:erg_IN1} with $F_{\gamma_{\mathrm{e2e}}}(t) = P_{\mathrm{s,out}}(t)$. Here $P_{\mathrm{s,out}}$ is given by \eqref{eq:ST_EH_out}.
\begin{remark}
For the interference dominant scenario, we can obtain analytical expressions for the secondary outage probability and the ergodic capacity by setting noise powers $N_{0,\mathrm{r}}$ and $N_{0,\mathrm{d}}$ to zero.
\end{remark}

\section{Results and Discussions}
\label{sec:results} 
\subsection{System Parameters and Simulation Setup}
Unless otherwise stated, we consider the following system parameters: The desired primary rate, $R_{\mathrm{p}} = \mathrm{0.1}\,\mathrm{bits/s/Hz}$, the desired secondary rate, $R_{\mathrm{s}} = \mathrm{0.1}\,\mathrm{bits/s/Hz}$, the primary outage threshold, $\Theta_{\mathrm{p}} = 0.1$, the energy conversion efficiency factor, $\delta = \mathrm{0.3}$, the primary transmit power, $P_{\mathrm{p}} = 10~\mathrm{dB}$, energy harvesting circuitry activation threshold, $P_{\mathrm{th}} = -10$~$\mathrm{dBm}$, bandwidth, $B = 1$~$\mathrm{Hz}$, noise powers, $N_{\mathrm{0,r}} = N_{\mathrm{0,d}} = N_0 = 0$~$\mathrm{dBm}$, and peak powers, $P_{\mathrm{t,ST}} = P_{\mathrm{t,SR}} = P_{\mathrm{t}} = 10$~$\mathrm{dB}$. We consider a 2-D simulation setup where ($x_{i}$, $y_{i}$) denotes the coordinate of $i$th user. The mean channel gain between $i$th and $j$th users is $d_{ij}^{-\rho}$, where $d_{ij}$ is the distance between users $i$ and $j$, and $\rho = 2.7$ is the path loss exponent.\footnote{Without loss of generality, we assume that the channel power gains (without path loss) are exponentially distributed with mean $1$.} The ST, SR, and SD are placed at ($\mathrm{0}$, $\mathrm{0}$), ($\mathrm{2}$, $\mathrm{0}$), and ($\mathrm{4}$, $\mathrm{0}$), respectively. The PTs are co-located at ($\mathrm{0}$, $\mathrm{2}$), while PDs are co-located at ($\mathrm{4}$, $\mathrm{2}$). 

\begin{figure}
\centering
\includegraphics[scale=0.43]{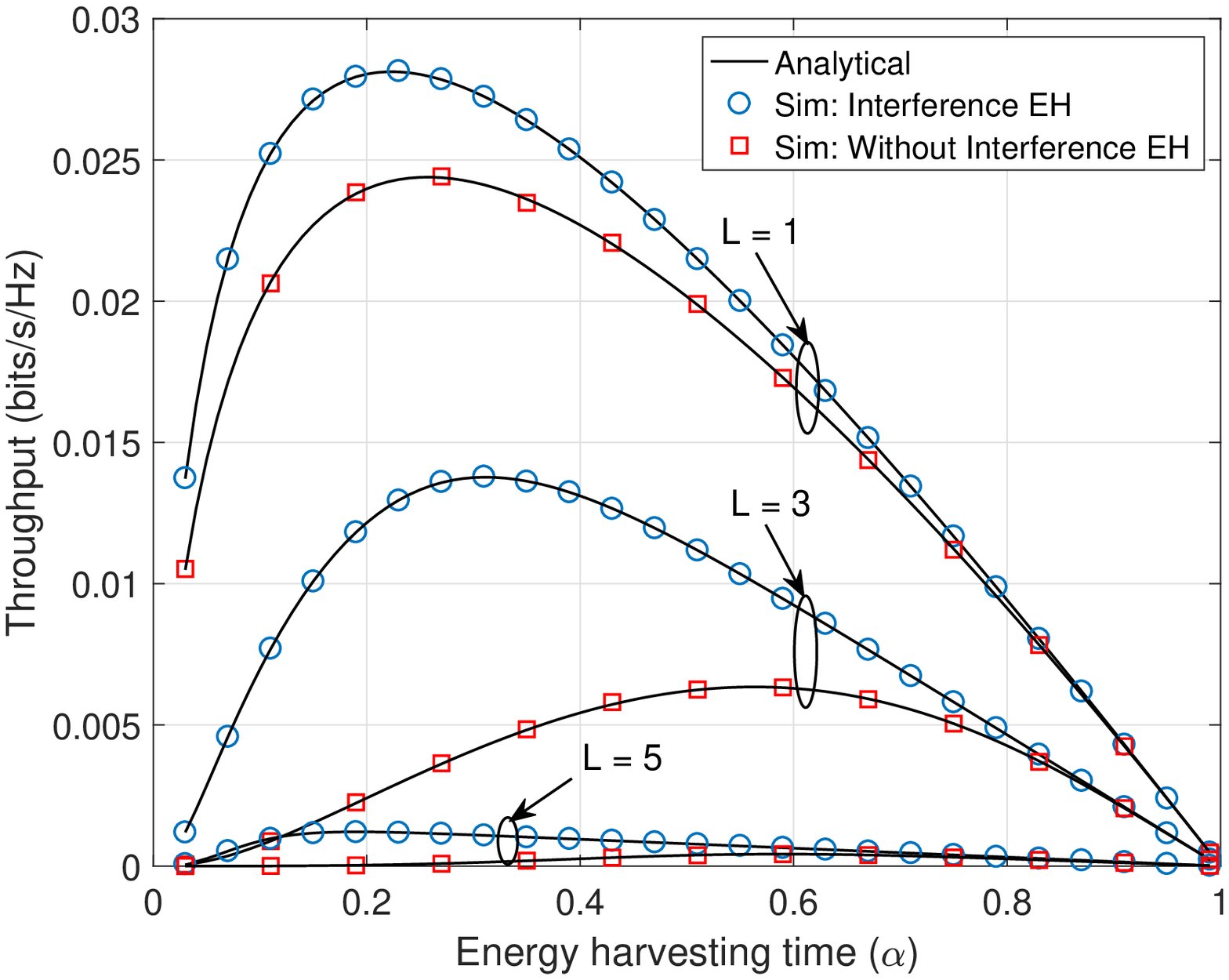}
\caption{Delay-limited transmission with non-EH ST: Interference EH versus without interference EH for different number of primary transceivers ($L$).}
\label{fig:int_no_int_delay_limited}
\end{figure}
\begin{figure}
\centering
\includegraphics[scale=0.43]{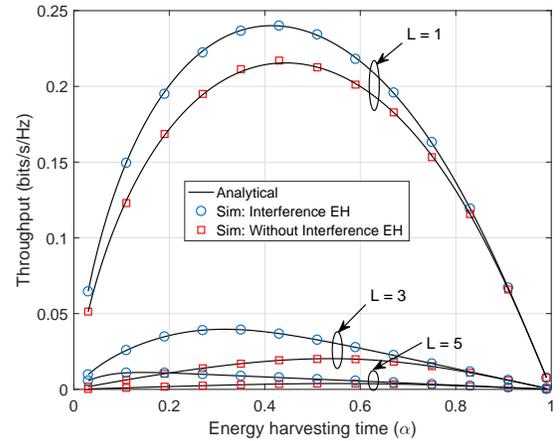}
\caption{Delay-tolerant transmission with non-EH ST: Interference EH versus without interference EH for different number of primary transceivers ($L$).}
\label{fig:int_no_int_delay_tolerant}
\end{figure}

\subsection{Effect of Interference-Aided Energy Harvesting\protect\footnote{The discussion in this subsection is for the interference plus noise scenario.}}
\label{sec:eff_L}
For ST as a non-EH node, Fig.~\ref{fig:int_no_int_delay_limited} plots the SU throughput for the delay-limited transmission against the energy harvesting ratio $\alpha$. As expected, the SR harvesting energy from the primary interference in addition to that from received information signals achieves a higher throughput than the conventional method where SR treats interference as an unwanted signal in EH phase. For example, for $L = 1$ and $\alpha = 0.4$, the gain in the secondary throughput due to energy harvesting from the primary interference is almost $10\%$ (see the curve corresponding to $L = 1$ in Fig.~\ref{fig:int_no_int_delay_limited}). For a given number of primary transceivers $L$, as $\alpha$ increases from 0 to 1, the SU throughput increases first and then decreases beyond the optimal value of $\alpha$ that maximizes the throughput. This trade-off can be attributed to two conflicting effects that are dependent on $\alpha$. The increase in $\alpha$ allows SR to harvest more energy, which improves the quality of the transmission on SR-SD link and thus increases the SU throughput. On the contrary, the time for information transmission reduces with the increase in $\alpha$, which decreases the SU throughput. A similar behavior of SU throughput against $\alpha$ can be observed for the delay-tolerant transmission with non-EH ST (see Fig.~\ref{fig:int_no_int_delay_tolerant}).

Though the increase in $L$ provides SR higher harvested energy, the interference to the information reception at both SR and SD increases, which decreases the SU throughput. Another negative consequence of an increase in $L$ is stricter primary outage constraints. Since SU should satisfy the outage constraint at each PD, the increase in the number of PUs makes the constraints more difficult to satisfy, which reduces the maximum allowed transmit powers for both ST and SR.

\begin{figure}
\centering
\includegraphics[scale=0.43]{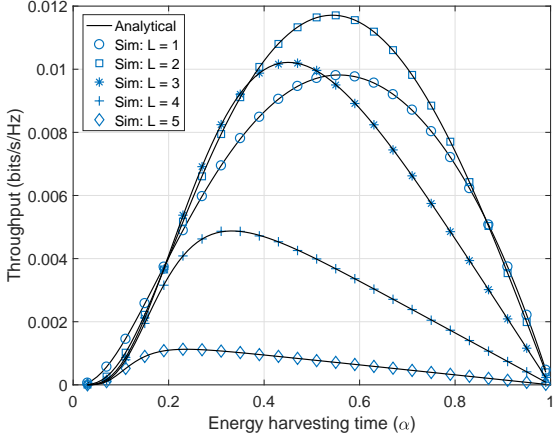}
\caption{Delay-limited transmission with EH ST: Interference EH versus without interference EH for different number of primary transceivers ($L$).}
\label{fig:int_no_int_delay_limited_EH_ST}
\end{figure}
Fig.~\ref{fig:int_no_int_delay_limited_EH_ST} plots the SU throughput versus $\alpha$ for the delay-limited transmission when both ST and SR are EH nodes. The SU throughput varies with $\alpha$ similar to that in Fig.~\ref{fig:int_no_int_delay_limited} (where ST is a non-EH node). But unlike the non-EH ST case, the SU throughput does not decrease monotonically with $L$ (See curves corresponding to $L = 1$ and $L = 2$ in Fig.~\ref{fig:int_no_int_delay_limited_EH_ST}.). This is because both ST and SR are dependent on primary signals to get the energy for transmissions. Thus, having more PTs is better to gather more energy. But, eventually, with an increase in $L$, the deteriorating effects of primary interference and primary outage constraints dominate the gain obtained due to higher harvested energy, and the SU throughput decreases. Also the secondary network with EH ST achieves a smaller throughput compared to that with non-EH ST as the former is more energy-constrained due to the absence of a conventional and reliable energy source like the latter.

\subsection{Effect of the Primary Outage Constraint}
\begin{figure}
\centering
\includegraphics[scale=0.43]{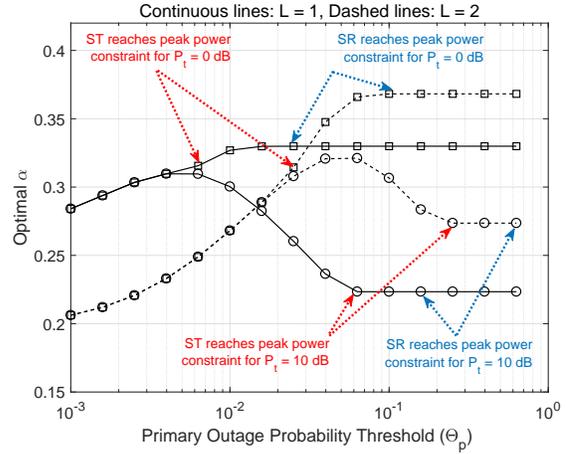}
\caption{Delay-limited transmission with non-EH ST: Optimal energy harvesting time ($\alpha$) versus the primary outage probability threshold ($\Theta_{\mathrm{p}}$). Curves with squares correspond to $P_{\rm{t}} = 0~\mathrm{dB}$, while curves with circles correspond to $P_{\rm{t}} = 10~\mathrm{dB}$.}
\label{fig:opt_alpha_delay_limited}
\end{figure}

\begin{figure}
\includegraphics[scale=0.43]{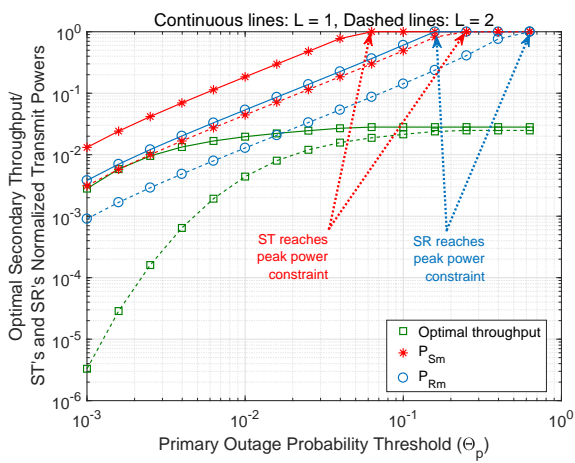}
\caption{Delay-limited transmission non-EH ST: Optimal throughput versus $\Theta_{\mathrm{p}}$, $P_{\mathrm{t}} = 10~\mathrm{dB}$.}
\label{fig:opt_through_delay_limited_pt_8}
\end{figure}

For the delay-limited transmission with non-EH ST, Fig.~\ref{fig:opt_alpha_delay_limited} shows the effect of the primary outage threshold ($\Theta_{\mathrm{p}}$) on the optimal $\alpha$ for different values of $L$ and peak power constraint of $P_{\mathrm{t}}$ in interference plus noise scenario. For $P_{\mathrm{t}} = 0~\mathrm{dB}$ and $P_{\mathrm{t}} = 10~\mathrm{dB}$, Figs.~\ref{fig:opt_through_delay_limited_pt_2} and \ref{fig:opt_through_delay_limited_pt_8}, respectively, plot the optimal SU throughput and the maximum allowed
transmit powers for ST and SR normalized by $P_{\mathrm{t}}$. From \eqref{eq:thro_IN} and \eqref{eq:Psout_IN}, we can see that, obtaining an analytical expression for the optimal $\alpha$ is difficult due to the involvement of an integral form of analytical expressions; but the optimal $\alpha$ (and hence the optimal SU throughput) can be obtained numerically.

\subsubsection{ For $P_{\mathrm{t}} = \mathrm{0}~\mathrm{dB}$}
\begin{figure}
\centering
\includegraphics[scale=0.43]{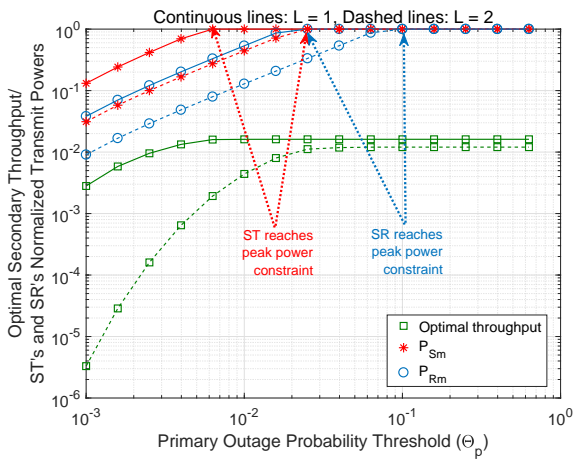}
\caption{Delay-limited transmission non-EH ST: Optimal throughput versus $\Theta_{\mathrm{p}}$, $P_{\mathrm{t}} = 0~\mathrm{dB}$.}
\label{fig:opt_through_delay_limited_pt_2}
\end{figure}

Fig.~\ref{fig:opt_alpha_delay_limited} shows that the optimal $\alpha$ increases with $\Theta_{\mathrm{p}}$ because the increase in $\Theta_{\mathrm{p}}$ allows ST and SR to transmit at higher powers. Hence $\alpha$ increases to cater relay's higher transmit power. Also, higher transmit powers of ST and SR increase SINRs at both SR and SD, respectively, which provide an extra margin to increase $\alpha$. The increased SINR overcomes the loss in information transmission time, improving SU throughput (see Fig.~\ref{fig:opt_through_delay_limited_pt_2}). The peak power constraints become active due to the increased maximum allowed powers at ST ($P_{\mathrm{ST}}$ given by \eqref{eq:PST}) and SR ($P_{\mathrm{SR}}$ given by \eqref{eq:Ppout7}) with an increase in $\Theta_{\mathrm{p}}$ beyond a threshold.
Figs.~\ref{fig:opt_alpha_delay_limited} and \ref{fig:opt_through_delay_limited_pt_2} show this behavior where we observe that ST reaches its peak power constraint first{\footnote{In the simulation setup, ST is located farther from primary destinations than SR. This allows ST to transmit at higher power than that of SR for the same $\Theta_{\mathrm{p}}$, which causes ST to reach the peak power constraint before SR.}} at $\Theta_{\mathrm{p}} = 6 \times 10^{-3}$ for $L = 1$ and $\Theta_{\mathrm{p}} = 2.5 \times 10^{-2}$ for $L = 2$, which forces ST to transmit with peak power $P_{\mathrm{t}}$ even though the further increase in $\Theta_{\mathrm{p}}$ allows it to transmit at higher power. After this point, to serve the increasing SR's transmit power, the optimal $\alpha$ increases at a faster rate than that without the peak power constraint until the peak power constraint at SR is reached. Once SR's peak power constraint is reached, SR is also forced to transmit at fixed power $P_{\mathrm{t}}$ for any further increase in $\Theta_{\mathrm{p}}$, and the optimal $\alpha$ remains the same thereafter (i.e., after $\Theta_{\mathrm{p}} = 2.5 \times 10^{-2}$ for $L = 1$).

\subsubsection{ For $P_{\mathrm{t}} = \mathrm{10}~\mathrm{dB}$}
As Fig.~\ref{fig:opt_alpha_delay_limited} shows, unlike for $P_{\mathrm{t}} = \mathrm{0}~\mathrm{dB}$, the optimal $\alpha$ does not increase monotonically with $\Theta_{\mathrm{p}}$; instead, it initially increases and then decreases after the tipping point. This is because the maximum allowed transmit power at ST increases faster than that at SR, and at the tipping point, before reaching the peak power constraint of $10~\mathrm{dB}$, it increases to the level that is more than sufficient to cater the increase in SR's transmit power. At this point, the secondary network is better off by allocating more time to information transmission. This behavior can be verified from Fig.~\ref{fig:opt_through_delay_limited_pt_8}, where the optimal SU throughput increases when the optimal $\alpha$ decreases. We do not observe this trend for $P_{\mathrm{t}} = $ $0~\mathrm{dB}$, as the peak power constraint becomes active before the arrival of the tipping point. For $P_{\mathrm{t}} = $ $10~\mathrm{dB}$, as ST reaches its peak power constraint (after $\Theta_{\mathrm{p}} = 6.3  \times 10^{-2}$ for $L = 1$ and $\Theta_{\mathrm{p}} = 2.5 \times 10^{-1}$ for $L = 2$), the optimal $\alpha$ remains the same as ST starts transmitting with constant power which is still sufficient to meet SR's power requirements. 

As discussed in Section \ref{sec:eff_L}, an increase in $L$ reduces the maximum allowed power at both ST and SR, which delays the arrival of the peak power constraint as shown in Figs.~\ref{fig:opt_through_delay_limited_pt_2} and \ref{fig:opt_through_delay_limited_pt_8} and the tipping point for $P_{\mathrm{t}} = $ $10~\mathrm{dB}$ (see Fig.~\ref{fig:opt_alpha_delay_limited}).

\begin{figure}
\centering
\includegraphics[scale=0.43]{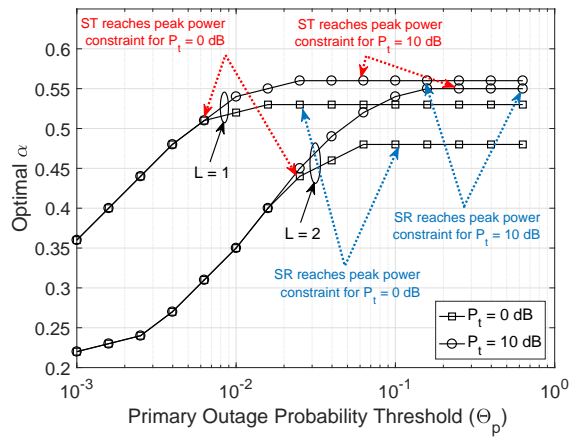}
\caption{Delay-limited transmission EH ST: Optimal energy harvesting time ($\alpha$) versus $\Theta_{\mathrm{p}}$.}
\label{fig:opt_alpha_EH_ST}
\end{figure}
\begin{figure}
\includegraphics[scale=0.43]{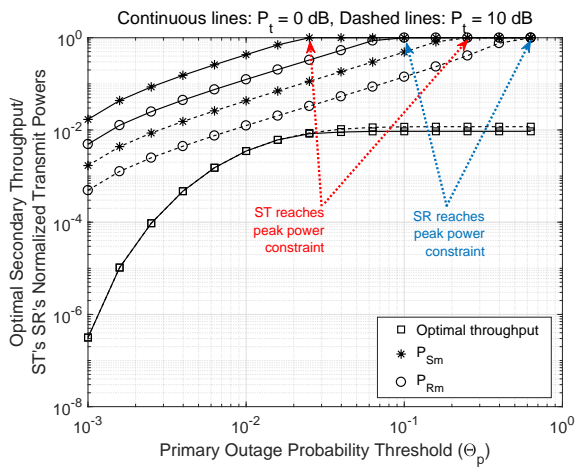}
	\caption{Delay-limited transmission EH ST: Optimal energy harvesting time ($\alpha$) versus $\Theta_{\mathrm{p}}$, \underline{$L = 2$}.}
	\label{fig:opt_alpha_EH_ST1}
\end{figure}

\subsubsection{EH secondary transmitter}
\label{sec:com_EHN}
For the case of EH ST, Fig.~\ref{fig:opt_alpha_EH_ST} shows that the optimal $\alpha$ behavior does not observe the tipping point for $P_{\mathrm{t}} = $ $10~\mathrm{dB}$ as that for non-EH ST case (see Fig.~\ref{fig:opt_alpha_delay_limited}); instead, the optimal $\alpha$ keeps increasing until ST reaches its peak power constraint. Also, unlike for the non-EH ST case (see Fig.~\ref{fig:opt_alpha_delay_limited}), the optimal $\alpha$ is higher for $P_{\mathrm{t}} = 10~\mathrm{dB}$ than that for $P_{\mathrm{t}} = 0~\mathrm{dB}$ after ST reaches its peak power constraint in the case of $P_{\mathrm{t}} = 0~\mathrm{dB}$. This behavior is due to the dependence of both ST and SR on primary signals to harvest energy in the case of EH ST. For EH ST case with $P_{\mathrm{t}} = 0~\mathrm{dB}$, as ST reaches the peak power constraint, it does not need to harvest any more energy and the rate of increase in the optimal $\alpha$ reduces as only SR harvests energy until its peak power constraint is reached, after which the optimal $\alpha$ remains the same (see Figs.~\ref{fig:opt_alpha_EH_ST} and \ref{fig:opt_alpha_EH_ST1}). On the other hand, for $P_{\mathrm{t}} = 10~\mathrm{dB}$, the peak power constraint for ST arrives later than that for  $P_{\mathrm{t}} = 0~\mathrm{dB}$. Hence the optimal $\alpha$ for $P_{\mathrm{t}} = 10~\mathrm{dB}$ is higher than that for $P_{\mathrm{t}} = 0~\mathrm{dB}$ to cater the energy requirements of ST and SR after the point where ST has reached its peak power constraint for the case of $P_{\mathrm{t}} = 0~\mathrm{dB}$. Unlike for the case of $P_{\mathrm{t}} = 0~\mathrm{dB}$, the optimal $\alpha$ remains constant with an increase in $\Theta_{\mathrm{p}}$ once ST reaches its peak power constraint, as the increase in $\alpha$ further causes no increase in ST's transmit power. Also, since ST transmits at a higher power than that at SR (see the footnote~6), the duration of energy harvesting is mainly impacted by the ST's energy requirement. By the time ST reaches its peak power constraint in the case of $P_{\mathrm{t}} = 10~\mathrm{dB}$, SR has already harvested enough energy needed for its transmission. Hence the secondary network can achieve a better performance by not further increasing the energy harvesting time but by allocating the time for information transmission.
	
From Figs. \ref{fig:opt_alpha_delay_limited} and \ref{fig:opt_alpha_EH_ST}, we observe that the optimal $\alpha$ is higher for EH ST case than that for non-EH ST case. This behavior is expected as, in EH ST case, both ST and SR depend on the primary interference to harvest energy; while, in non-EH ST case, only SR is an EH node.

\begin{figure}
	\centering
\includegraphics[scale = 0.43]{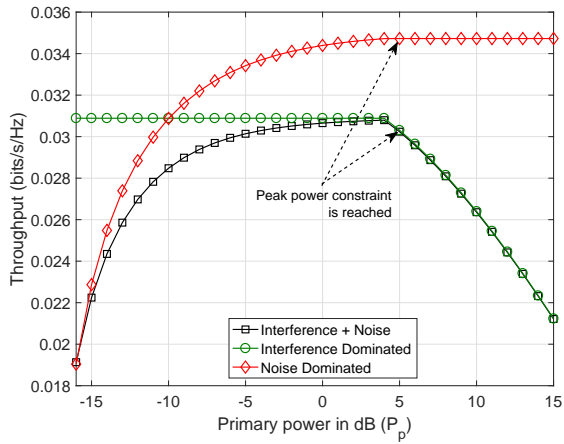}
	\caption{Effect of primary's transmit power on SU's throughput performance, PTs' location: (1,3), $L = 1$, $\alpha = 0.3$.}
	\label{fig:int_noise}
\end{figure}

\subsection{Effect of Primary's Transmit Power}
When ST is a non-EH node, Fig.~\ref{fig:int_noise} shows the effect of primary's transmit power $P_{\mathrm{p}}$ on SU throughput for a fixed $\alpha$. For small values of $P_{\mathrm{p}}$ (below $-15$ $\mathrm{dB}$), the noise dominates the SU throughput performance, and the curve corresponding to the interference plus noise case matches that corresponding to the noise dominant case. On the other hand, at higher values of $P_{\mathrm{p}}$ (above $4$ $\mathrm{dB}$), the interference dominates the SU throughput performance, and the curve corresponding to the interference plus noise case matches that corresponding to the interference dominant case. An increase in $P_{\mathrm{p}}$ increases the energy harvested at SR and the signal strength at primary destinations, which allows an increase in the transmit powers of ST and SR. These positive effects dominate the increased primary interference at SR and SD until ST reaches the peak power constraint, after which ST transmits with constant power irrespective of a further increase in $P_{\mathrm{p}}$. Hence the SINRs at SR and SD reduces and the SU throughput decreases. For the noise dominant case, the throughput remains almost constant after ST reaches the peak power constraint, as the harvested energy at SR remains the same, thereby its transmit power.
 
\section{Conclusions and Future Works}
\label{sec:conc}
We considered a scenario of spectrum sharing between a secondary network and multiple primary transceivers, where secondary users communicate via an interference-aided energy harvesting decode-and-forward relay under primary outage constraints. For both cases of energy harvesting and non-energy harvesting secondary transmitters, we derived exact analytical expressions for the secondary outage probability and the ergodic capacity in delay-limited and delay-tolerant transmission modes, respectively. For the non-energy harvesting secondary transmitter case, the results show that, though the primary interference serves an extra energy source, the deteriorating effects of the interference dominate the SU throughput. On the contrary, for the energy harvesting secondary transmitter case, the gain obtained by harvesting energy from primary interference could dominate the detrimental effects of the primary interference. We observe through numerical results that the peak power constraints at the secondary transmitter and the secondary relay play an important role in deciding the optimal energy harvesting time.

An interesting future direction is the use of mathematical tools from stochastic geometry to analyze the performance of the secondary network where the primary transceivers are randomly located. Also, the use of multiple antennas at the secondary relay can help harvest higher amount of energy and mitigate the detrimental effects of the primary interference in a better manner.

\appendices
\section{Proof of \eqref{eq:PST}}
\label{sec:PST}
Let $\mathcal{K}$ be $\mathbb{P}\left( \frac{P_{\mathrm{p}}|h_{i}|^{2}}{P_{\mathrm{ST}}|h_{\mathrm{s}i}|^{2} + N_0} \leq \zeta_{\mathrm{p}}\right) $. Then we can write
\begin{align}
\mathcal{K} = \int_{0}^{\infty} \mathbb{P}\left( \frac{P_{\mathrm{p}}|h_{i}|^{2}}{P_{\mathrm{ST}}y + N_0} \leq \zeta_{\mathrm{p}}\right)f_{|h_{\mathrm{s}i}|^{2}}(y)\mathrm{d}y,
\label{eq:Ppout3}
\end{align}

\noindent where $f_{|h_{\mathrm{s}i}|^{2}}(y)$ is the probability density function (PDF) of $|h_{\mathrm{s}i}|^{2}$, which is given by $f_{|h_{\mathrm{s}i}|^{2}}(y) = \frac{1}{\lambda_{\mathrm{sp}}}\exp\left(-\frac{y}{\lambda_{\mathrm{sp}}}\right)$. Solving \eqref{eq:Ppout3} and then substituting the value of $\mathcal{K}$ in \eqref{eq:Ppout2}, we obtain 
\begin{align}
\mathrm{P}_{\mathrm{p, out, ST}}  = 1 - \left(\frac{\mathcal{A}P_{\mathrm{p}}\lambda_{\mathrm{pp}}}{P_{\mathrm{ST}}\lambda_{\mathrm{sp}}\zeta_{\mathrm{p}} + P_\mathrm{PT}\lambda_{\mathrm{pp}}}\right)^{L},
\label{eq:Ppout5}
\end{align}
where $\mathcal{A} = \exp\left(-\frac{\zeta_{\mathrm{p}}N_0 B}{L P_{\mathrm{p}}\lambda_{\mathrm{pp}}}\right)$.
Solving \eqref{eq:Ppout5} for $P_{\mathrm{ST}}$, we obtain the required expression in \eqref{eq:PST}.

\section{Proof of Proposition~\ref{prop:power_outage_start}}
\label{app:power_outage_start}
Given ST transmits with power $P_{\mathrm{Sm}}$ and each PT transmits with power $P_{\mathrm{p}}$, the received power at SR is 
\begin{equation}
P_{\mathrm{rec}} = P_{\mathrm{Sm}}|h_{\mathrm{sr}}|^{2} + \sum_{i = 1}^{L} P_{\mathrm{p}}|h_{i\mathrm{r}}|^2.
\label{eq:peout1}
\end{equation}
The power outage probability follows as
\begin{align}
P_{\mathrm{e,out}} = \mathbb{P}(P_{\mathrm{rec}} < P_{\mathrm{th}}).
\label{eq:peout2}
\end{align}

\textit{If} $P_\mathrm{p}\lambda_{\mathrm{pr}} \neq P_{\mathrm{Sm}}\lambda_{\mathrm{sr}}$: Let $X = P_{\mathrm{Sm}}|h_{\mathrm{sr}}|^{2}$ and $Y = \sum_{i = 1}^{L} P_{\mathrm{p}}|h_{i\mathrm{r}}|^2$. Then $X$ and $Y$ are exponential and gamma RVs, respectively. The PDFs of $X$ and $Y$ are given by
\begin{align}
f_X(x) = \frac{1}{P_{\mathrm{Sm}}\lambda_{\mathrm{sr}}}\exp\left(-\frac{x}{P_{\mathrm{Sm}}\lambda_{\mathrm{sr}}}\right), \quad x \geq 0,
\end{align}
and 
\begin{align}
f_Y(y) = \frac{y^{L-1}\exp\left(-\frac{y}{P_{\mathrm{p}}\lambda_{\mathrm{pr}}}\right)}{\Gamma(L)(P_{\mathrm{p}}\lambda_{\mathrm{pr}})^{L}}, \quad y \geq 0,
\end{align}
respectively. From \eqref{eq:peout1}, we can write
\begin{align}
P_{\mathrm{e,out}} &= \mathbb{P}(X + Y < P_{\mathrm{th}}) \nonumber \\
& = \int_{y = 0}^{P_{\mathrm{th}}} \int_{x = 0}^{P_{\mathrm{th}}-y} f_{X}(x)f_Y(y)\mathrm{d}x\,\mathrm{d}y  \nonumber \\
&= \int_{y = 0}^{P_{\mathrm{th}}} \left[1 - \exp\left(-\frac{P_{\mathrm{th}}-y}{P_{\mathrm{Sm}}\lambda_{\mathrm{sr}}}\right)\right]\frac{y^{L-1}\exp\left(-\frac{y}{\lambda_{\mathrm{pr}}}\right)}{\Gamma(L)(P_{\mathrm{p}}\lambda_{\mathrm{pr}})^{L}}\mathrm{d}y \nonumber\\
& = \int_{y = 0}^{P_{\mathrm{th}}} \frac{y^{L-1}\exp\left(-\frac{y}{\lambda_{\mathrm{pr}}}\right)}{\Gamma(L)(P_{\mathrm{p}}\lambda_{\mathrm{pr}})^{L}}\mathrm{d}y \nonumber \\
&- \frac{\exp\left(-\frac{P_{\mathrm{th}}}{P_{\mathrm{Sm}\lambda_{\mathrm{sr}}}}\right)}{\Gamma(L)(P_{\mathrm{p}}\lambda_{\mathrm{pr}})^{L}}\int_{y = 0}^{P_{\mathrm{th}}} y^{L-1}\exp\left(-\omega y\right)\mathrm{d}y.
\label{eq:peout3}
\end{align}
Solving \eqref{eq:peout3}, we get the required expression of the power outage probability for $P_\mathrm{p}\lambda_{\mathrm{pr}} \neq P_{\mathrm{Sm}}\lambda_{\mathrm{sr}}$.

\textit{If} $P_\mathrm{p}\lambda_{\mathrm{pr}} = P_{\mathrm{Sm}}\lambda_{\mathrm{sr}}$: $X + Y$ is the sum of $L+1$ i.i.d. exponential RVs. Letting $Z = X + Y$, $P_{\mathrm{e,out}}$ is the CDF of $Z$, where $Z$ is a gamma RV with shape parameter $L+1$ and scale parameter ${P_{\mathrm{p}}\lambda_\mathrm{pr}}$.

\section{Proof of Proposition \ref{prop:psout_IN}}
\label{sec:psout_IN}
From \eqref{eq:Psout_IN_basic}, we have 
\begin{align}
P_{\mathrm{s,out}} = 1 - \mathbb{P}\left(\gamma_{\mathrm{SR}} \geq \zeta_{\mathrm{s}}, \gamma_{\mathrm{SD}} \geq \zeta_{\mathrm{s}}\right),
\end{align}
where 
\begin{equation}
\gamma_{\mathrm{SR}} = \frac{X}{U + N_{0}},
\label{eq:SINR_SR_app}
\end{equation}

\begin{align}
\gamma_{\mathrm{SD}} = \frac{\min\left( \theta \left(X + U \right), P_{\mathrm{Rm}}\right) Y}{Z + N_{0}},
\label{eq:SINR_SD_app}
\end{align}
with $\theta = \frac{2 \alpha \delta}{1-\alpha}$, $X = P_{\mathrm{Sm}}|h_{\mathrm{sr}}|^{2}$, $Y = |h_{\mathrm{rd}}|^{2}$, $Z = \sum_{i = 1}^{L}P_{\mathrm{p}}|h_{i\mathrm{d}}|^2$, and $U = \sum_{i = 1}^{L}P_{\mathrm{p}}|h_{i\mathrm{r}}|^2$. $X$ and $Y$ are exponentially distributed with means $\lambda_{\mathrm{x}} = P_{\mathrm{Sm}}\lambda_{\mathrm{sr}}$ and $\lambda_{\mathrm{y}} = \lambda_{\mathrm{rd}}$, respectively, while $Z$ and $U$ are gamma RVs with shape parameter $L$ and scale parameters $P_{\mathrm{p}}\lambda_{\mathrm{pd}}$ and $P_{\mathrm{p}}\lambda_{\mathrm{pr}}$, respectively. Note that, $X$, $Y$, $Z$, and $U$ are mutually independent RVs, but $\gamma_{\mathrm{SR}}$ and $\gamma_{\mathrm{SD}}$ are dependent RVs due to the presence of the common RVs $X$ and $U$. Then we can express the secondary outage probability as 
\begin{align}
P_{\mathrm{s,out}} &= 1 - \mathcal{I},
\label{eq:long_out}
\end{align}
where
\vspace*{-4mm}

{{\small
\begin{align*}
\!\mathcal{I} &=\!\!\!\!\int\limits_{u = 0}^{\infty}\int\limits_{x = \zeta_{\mathrm{s}}(u+N_0)}^{\infty}\int\limits_{z = 0}^{\infty}\int\limits_{y = \frac{\zeta_{\mathrm{s}}(z+N_0)}{\min(\theta(x + u), P_{\mathrm{Rm}})}}^{\infty}\hspace{-12mm}f_U(u)f_X(x)f_Z(z)f_Y(y)\mathrm{d}u\mathrm{d}x\mathrm{d}z\mathrm{d}y
\end{align*}}}\vspace*{-3mm}

\noindent with $f_U(u)$, $f_X(x)$, $f_Z(z)$, and $f_Y(y)$ are PDFs of $U$, $X$, $Z$, and $Y$, respectively. Integrating over $Y$, we obtain
\begin{align}
\mathcal{I} &=\int\limits_{u = 0}^{\infty}\int\limits_{x = \zeta_{\mathrm{s}}(u+N_0)}^{\infty}\int\limits_{z = 0}^{\infty}f_U(u)f_X(x)f_Z(z) \nonumber \\
& \times \exp\left(-\frac{\zeta_{\mathrm{s}}(z+N_0)}{\min(\theta(x + u), P_{\mathrm{Rm}})\lambda_{\mathrm{y}}}\right)\mathrm{d}u \, \mathrm{d}x \,\mathrm{d}z.
\label{eq:I2}
\end{align}
Now, let 
\begin{align}
\mathcal{I}_1 &= \int\limits_{z = 0}^{\infty}f_Z(z)\exp\left(-\frac{\zeta_{\mathrm{s}}(z+N_0)}{\min(\theta(x + u), P_{\mathrm{Rm}})\lambda_{\mathrm{y}}}\right)\mathrm{d}z.
\label{eq:I3}
\end{align}
Given that $Z$ is a gamma RV and letting $\mathcal{A}(u,x) = \min(\theta(x +u),P_{\mathrm{Rm}})\lambda_{\mathrm{y}}$, we can simplify \eqref{eq:I3} as
\begin{align}
\mathcal{I}_1 &= \frac{\exp\left(-\frac{\zeta_{\mathrm{s}}N_0}{\mathcal{A}(u,x)}\right)}{\Gamma(L)(P_{\mathrm{p}}\lambda_{\mathrm{z}})^{L}}\!\!\int\limits_{z = 0}^{\infty} \!\!z^{L-1}\exp \!\left(\!-\left(\!\frac{1}{P_{\mathrm{p}}\lambda_{\mathrm{z}}}+\frac{\zeta_{\mathrm{s}}}{\mathcal{A}(u,x)}\!\right)\! z \!\right)\!\mathrm{d}z,
\label{eq:I5}
\end{align}
where $\lambda_{\mathrm{z}} = \lambda_{\mathrm{pd}}$. Using the definition of the gamma function as $\Gamma(L) = \int_{t = 0}^{\infty} t^{L-1}\exp(-t)\mathrm{d}t$, we can finally write \eqref{eq:I5} as
\begin{align}
\mathcal{I}_1 &= \frac{\exp\left(-\frac{\zeta_{\mathrm{s}}N_0}{\mathcal{A}(u,x)}\right)}{(P_{\mathrm{p}}\lambda_{\mathrm{z}})^{L}}\left[\frac{1}{P_{\mathrm{p}}\lambda_{\mathrm{z}}}+\frac{\zeta_{\mathrm{s}}}{\mathcal{A}(u,x)}\right]^{-L}.
\label{eq:I6}
\end{align}
Using \eqref{eq:I3} and \eqref{eq:I6}, we can write \eqref{eq:I2} as
\begin{align}
\mathcal{I} &=\int\limits_{u = 0}^{\infty}\int\limits_{x = \zeta_{\mathrm{s}}(u+N_0)}^{\infty}\frac{\exp\left(-\frac{\zeta_{\mathrm{s}}N_0}{\mathcal{A}(u,x)}\right)}{(P_{\mathrm{p}}\lambda_{\mathrm{z}})^{L}}\left[\frac{1}{P_{\mathrm{p}}\lambda_{\mathrm{z}}}+\frac{\zeta_{\mathrm{s}}}{\mathcal{A}(u,x)}\right]^{-L} \nonumber \\
& \times f_U(u)f_X(x) \mathrm{d}u \, \mathrm{d}x.
\label{eq:I7}
\end{align}
Given that $\mathcal{A}(u,x)$ is the minimum of two terms, we can split it as
\begin{equation}
\label{eq:binary12}
\mathcal{A}(u,x) = \left\{
  \begin{array}{l l}
    \theta(x +u)\lambda_{\mathrm{y}}, & \quad \text{if}~x \leq \frac{P_{\mathrm{Rm}} -\theta u}{\theta} \\
    P_{\mathrm{Rm}}\lambda_{\mathrm{y}}, & \quad \text{if}~x > \frac{P_{\mathrm{Rm}} -\theta u}{\theta}.\\
  \end{array} \right.
\end{equation}
Comparing the threshold $\frac{P_{\mathrm{Rm}} -\theta u}{\theta}$ for $x$ given in \eqref{eq:binary12} with the lower limit of $x$ in \eqref{eq:I7}, we can split \eqref{eq:I7} as
\begin{equation}
\mathcal{I} = \frac{1}{(P_{\mathrm{p}}\lambda_{\mathrm{z}})^{L}}(I_1 + I_2 + I_3),
\label{eq:I8}
\end{equation}
where
\begin{align}
I_1 &= \int\limits_{u = 0}^{\delta_1} \int\limits_{x = \zeta_{\mathrm{s}}(u+N_0)}^{\frac{P_{\mathrm{Rm}} -\theta u}{\theta}} f_U(u)f_X(x)\exp\left(-\frac{\zeta_{\mathrm{s}}N_0}{\theta(x +u)\lambda_{\mathrm{y}}}\right) \nonumber \\
& \times \left(\frac{1}{P_{\mathrm{p}}\lambda_{\mathrm{z}}}+\frac{\zeta_{\mathrm{s}}}{\theta(x +u)\lambda_{\mathrm{y}}}\right)^{-L} \mathrm{d}u\,\mathrm{d}x,
\label{eq:I9}
\end{align}

\begin{align}
I_2 &= \int\limits_{u = 0}^{\delta_1} \int\limits_{x = \frac{P_{\mathrm{Rm}} -\theta u}{\theta}}^{\infty} f_U(u)f_X(x)\exp\left(-\frac{\zeta_{\mathrm{s}}N_0}{P_{\mathrm{Rm}}\lambda_{\mathrm{y}}}\right) \nonumber \\
& \times \left(\frac{1}{P_{\mathrm{p}}\lambda_{\mathrm{z}}}+\frac{\zeta_{\mathrm{s}}}{P_{\mathrm{Rm}}\lambda_{\mathrm{y}}}\right)^{-L} \mathrm{d}u\,\mathrm{d}x,
\label{eq:I10}
\end{align}
and
\begin{align}
I_3 &= \int\limits_{u = \delta_1}^{\infty}\int\limits_{x = \zeta_{\mathrm{s}}(u+N_0)}^{\infty} f_U(u) f_X(x)\exp\left(-\frac{\zeta_{\mathrm{s}}N_0}{P_{\mathrm{Rm}}\lambda_{\mathrm{y}}}\right) \nonumber \\
&\times \left(\frac{1}{P_{\mathrm{p}}\lambda_{\mathrm{z}}}+\frac{\zeta_{\mathrm{s}}}{P_{\mathrm{Rm}}\lambda_{\mathrm{y}}}\right)^{-L} \mathrm{d}u\,\mathrm{d}x,
\label{eq:I11}
\end{align}
where $\delta_1 = \frac{P_{\mathrm{Rm}} - \zeta_{\mathrm{s}}\theta N_0}{(1+\zeta_{\mathrm{s}}\theta)}$. The double integral in \eqref{eq:I9} cannot be expressed in a closed form but it can be easily evaluated numerically. We can express the double integrals in \eqref{eq:I10} and \eqref{eq:I11} in closed forms as follows:
\begin{align}
I_2 &=\exp\left(-\frac{\zeta_{\mathrm{s}}N_0}{P_{\mathrm{Rm}}\lambda_{\mathrm{y}}}\right) \left(\frac{1}{P_{\mathrm{p}}\lambda_{\mathrm{z}}}+\frac{\zeta_{\mathrm{s}}}{P_{\mathrm{Rm}}\lambda_{\mathrm{y}}}\right)^{-L} \times  I_{21}
\label{eq:I12}
\end{align}
where 
\begin{align}
I_{21} =  \int\limits_{u = 0}^{\delta_1} \int\limits_{x = \frac{P_{\mathrm{Rm}} -\theta u}{\theta }}^{\infty} f_U(u)f_X(x)\mathrm{d}u\,\mathrm{d}x.
\label{eq:I13}
\end{align}
Averaging over $X$, we can write \eqref{eq:I13} as
\begin{align}
I_{21} =  \int\limits_{u = 0}^{\delta_1} \exp\left(- \frac{P_{\mathrm{Rm}} -\theta u}{\theta \lambda_{\mathrm{x}}}\right)\frac{u^{L-1}\exp\left(-\frac{u}{P_{\mathrm{p}}\lambda_{\mathrm{u}}}\right)}{\Gamma(L)(P_{\mathrm{p}}\lambda_{\mathrm{u}})^{L}}\mathrm{d}u.
\label{eq:I14}
\end{align}
Using the definition of lower incomplete gamma function as $\gamma(a,b) = \int_{0}^{b} t^{a-1}\exp(-t)\mathrm{d}t$, we can write \eqref{eq:I14} as
\begin{align}
I_{21} =  \frac{\gamma(L,\omega \delta_1) \exp\left(-\frac{P_{\mathrm{Rm}}}{\theta P_{\mathrm{Sm}}\lambda_{\mathrm{sr}}}\right)}{\Gamma(L)(P_{\mathrm{p}}\lambda_{\mathrm{pr}})^{L} \omega^{L}},
\label{eq:I15}
\end{align}
where $\omega$ is given by \eqref{eq:omegag}. Substituting \eqref{eq:I15} in \eqref{eq:I12}, we get a closed-form expression of $I_2$ as in \eqref{eq:I22}. Proceeding in the similar manner to obtain a closed-form for $I_2$, we can obtain a closed-form expression for $I_3$ as given in \eqref{eq:I33}. Once $I_1$, $I_2$, and $I_3$ are calculated, we can get $\mathcal{I}$ from \eqref{eq:I8}, and in turn, we can express the secondary outage probability as \eqref{eq:long_out}.

\section{Proof of Proposition \ref{prop:psout_N}}
\label{app:noise}

Let $\theta = \frac{2 \alpha \delta}{1-\alpha}$, $X = |h_{\mathrm{sr}}|^{2}$, and $Y = |h_{\mathrm{rd}}|^{2}$. Then $X$ and $Y$ are exponentially distributed RVs with means $\lambda_{\mathrm{x}} = \lambda_{\mathrm{sr}}$ and $\lambda_{\mathrm{y}} = \lambda_{\mathrm{rd}}$, respectively. Using \eqref{eq:SNR_SR_noise}, \eqref{eq:SNR_SD_noise}, and \eqref{eq:Psout_IN_basic}, we can write the secondary outage probability as
\begin{align}
P_{\mathrm{s,out}} &= 1 - \mathbb{P}\left(\frac{P_{\mathrm{Sm}}X}{N_0} \geq \zeta_{\mathrm{s}}, \frac{\min\left(\theta P_{\mathrm{Sm}}X,P_{\mathrm{Rm}}\right)Y}{N_0} \geq \zeta_{\mathrm{s}}\right) \nonumber \\
&= 1 - \int\limits_{x = \frac{\zeta_{\mathrm{s}}N_0}{P_{\mathrm{Sm}}}}^{\infty} \int\limits_{y = \frac{\zeta_{\mathrm{s}}N_0}{\min\left(\theta P_{\mathrm{Sm}}x,P_{\mathrm{Rm}}\right)}}^{\infty} f_X(x)f_Y(y)\mathrm{d}x\,\mathrm{d}y,
\label{eq:N1}
\end{align}
where $f_X(x)$ and $f_Y(y)$ are the PDFs of $X$ and $Y$, respectively. Following the procedure in Appendix \ref{sec:psout_IN}, we have following two cases based on $\min\left(\theta P_{\mathrm{Sm}}x,P_{\mathrm{Rm}}\right)$ and the lower limit of the integral corresponding to $X$, i.e., $x = \zeta_{\mathrm{s}}N_0$. 
\begin{itemize}
\item Case I: $\frac{P_{\mathrm{Rm}}}{\theta P_{\mathrm{Sm}}} \geq \frac{N_0 \zeta_{\mathrm{s}}}{P_{\mathrm{Sm}}}$. 
\item Case II: $\frac{P_{\mathrm{Rm}}}{\theta P_{\mathrm{Sm}}} < \frac{N_0 \zeta_{\mathrm{s}}}{P_{\mathrm{Sm}}}$.
\end{itemize}
We shall first derive $P_{\mathrm{s,out}}$ for Case I. \\

\noindent \textit{Case I}: We can split \eqref{eq:N1} as
\begin{align}
P_{\mathrm{s,out}} = 1 &- \left(\int\limits_{x = \frac{N_0 \zeta_{\mathrm{s}}}{P_{\mathrm{Sm}}}}^{\frac{P_{\mathrm{Rm}}}{\theta P_{\mathrm{Sm}}}} \int\limits_{y = \frac{N_0 \zeta_{\mathrm{s}}}{\theta P_{\mathrm{Sm}}x}}^{\infty} f_X(x)f_Y(y)\mathrm{d}x\,\mathrm{d}y \right.\nonumber \\
& + \left. \int\limits_{x = \frac{P_{\mathrm{Rm}}}{\theta P_{\mathrm{Sm}}}}^{\infty} \int\limits_{y = \frac{N_0 \zeta_{\mathrm{s}}}{P_{\mathrm{Rm}}}}^{\infty} f_X(x)f_Y(y)\mathrm{d}x\,\mathrm{d}y \right).
\label{eq:N2}
\end{align}
Simplifying \eqref{eq:N2}, we get
\begin{align}
P_{\mathrm{s,out}}& = 1 -\frac{1}{\lambda_{{x}}}\left(\int\limits_{x = \frac{N_0 \zeta_{\mathrm{s}}}{P_{\mathrm{Sm}}}}^{\frac{P_{\mathrm{Rm}}}{\theta P_{\mathrm{Sm}}}} \exp\left(-\frac{x}{\lambda_{\mathrm{x}}}\right)\exp\left(-\frac{N_0 \zeta_{\mathrm{s}}}{\theta P_{\mathrm{Sm}}x \lambda_{{y}}}\right) \right. \nonumber \\
& \left. + \int\limits_{x = \frac{P_{\mathrm{Rm}}}{\theta P_{\mathrm{Sm}}}}^{\infty} \exp\left(-\frac{x}{\lambda_{{\mathrm{x}}}}\right)\exp\left(-\frac{N_0 \zeta_{\mathrm{s}}}{P_{\mathrm{Rm}}
 \lambda_{{\mathrm{y}}}}\right)\right).
 \label{eq:N3}
\end{align}
Scaling the limits of the first integral in \eqref{eq:N3} appropriately and using the definition of the incomplete Bessel function~\cite{haaris} as $K_{\nu}(a,b) = \int_{1}^{\infty}\frac{\exp\left(-at - \frac{b}{t}\right)}{t^{\nu+1}} \mathrm{d}t$, we can express \eqref{eq:N3} as $I_4$, given in~\eqref{eq:I4N}. \\

\noindent \textit{Case II}: We can write~\eqref{eq:N1} as
\begin{align}
P_{\mathrm{s,out}} &= 1 - \int\limits_{x = \frac{\zeta_{\mathrm{s}}N_0}{P_{\mathrm{Sm}}}}^{\infty} \frac{1}{\lambda_{{\mathrm{x}}}}\exp\left(-\frac{x}{\lambda_{\mathrm{x}}}\right)\exp\left(-\frac{\zeta_{\mathrm{s}} N_0}{P_{\mathrm{Rm}}\lambda_{{y}}}\right)\mathrm{d}x.
\label{eq:N4}
\end{align}
Here, since $\frac{P_{\mathrm{Rm}}}{\theta P_{\mathrm{Sm}}} < \frac{N_0 \zeta_{\mathrm{s}}}{P_{\mathrm{Sm}}}$, we do not get the case of $\theta P_{\mathrm{Sm}} x < P_{\mathrm{Rm}}$ as the lower limit of the integral is greater than $\frac{P_{\mathrm{Rm}}}{\theta P_{\mathrm{Sm}}}$. Solving \eqref{eq:N4}, we get $I_5$ as in \eqref{eq:I5N}.

\section{Proof of Proposition \ref{prop:ST_EH_out}}
\label{app:ST_EH}
In addition to previously defined $\theta = \frac{2 \alpha \delta}{1-\alpha}$, $X = |h_{\mathrm{sr}}|^{2}$, $Y = |h_{\mathrm{rd}}|^{2}$, $Z = \sum_{i = 1}^{L}P_{\mathrm{p}}|h_{i\mathrm{d}}|^2$, and $U = \sum_{i = 1}^{L}P_{\mathrm{p}}|h_{i\mathrm{r}}|^2$, let us denote $V = \sum_{i = 1}^{L}P_{\mathrm{p}}|h_{i\mathrm{s}}|^2$. Then $V$ is a gamma RV with shape parameter $L$ and scale parameter $P_{\mathrm{p}}\lambda_{v}$ with $\lambda_{v} = \lambda_{\mathrm{ps}}$. From~\eqref{eq:Psout_IN_basic}, we can write the secondary outage probability as
\begin{equation}
P_{\mathrm{s,out}} = 1 - \mathbb{P}\left(\gamma_{\mathrm{SR}} \geq \zeta_{\mathrm{s}}, \gamma_{\mathrm{SD}} \geq \zeta_{\mathrm{s}}\right),
\label{eq:ST_EH_1}
\end{equation}
where $\gamma_{\mathrm{SR}}= \frac{\min\left(\theta V, P_{\mathrm{Sm}}\right) X}{U + N_{0,\mathrm{r}}}$ and $\gamma_{\mathrm{SD}}= \frac{\min\left(\theta U, P_{\mathrm{Rm}}\right) Y}{Z + N_{0,\mathrm{d}}}$. Note that $\gamma_{\mathrm{SR}}$ and $\gamma_{\mathrm{SD}}$ are dependent RVs due to the presence of the common RV $U$. To represent \eqref{eq:ST_EH_1} in terms of independent RVs, we condition $P_{\mathrm{s,out}}$ in \eqref{eq:ST_EH_1} on $U$. It follows that
\begin{align}
\mathbb{P}\left(\gamma_{\mathrm{SR}} \geq \zeta_{\mathrm{s}}, \gamma_{\mathrm{SD}} \geq \zeta_{\mathrm{s}}|U = u\right) &=  \mathbb{P}\left(\gamma_{\mathrm{SR}} \geq \zeta_{\mathrm{s}}|U = u \right) \nonumber \\
&\times  \mathbb{P}\left(\gamma_{\mathrm{SD}} \geq \zeta_{\mathrm{s}}|U = u \right),
\end{align}
where $\mathbb{P}\left(\gamma_{\mathrm{SR}} \geq \zeta_{\mathrm{s}}|U = u \right)$ and $\mathbb{P}\left(\gamma_{\mathrm{SD}} \geq \zeta_{\mathrm{s}}|U = u \right)$ are independent probabilities for a given $U = u$. We derive below $\mathbb{P}\left(\gamma_{\mathrm{SR}} \geq \zeta_{\mathrm{s}}|U = u \right)$ and $\mathbb{P}\left(\gamma_{\mathrm{SD}} \geq \zeta_{\mathrm{s}}|U = u \right)$. 

We can express $\mathbb{P}\left(\gamma_{\mathrm{SR}} \geq \zeta_{\mathrm{s}}|U = u \right)$ as
\begin{align}
&\mathbb{P}\left(\gamma_{\mathrm{SR}} \geq \zeta_{\mathrm{s}}|U = u \right) = \mathbb{P}\left(\frac{\min\left(\theta V, P_{\mathrm{Sm}}\right) X}{u + N_{0,\mathrm{r}}} \geq \zeta_{\mathrm{s}}\right) \nonumber \\
&= \int\limits_{v = 0}^{\infty} \int\limits_{x = \frac{(u+N_{0, \mathrm{r}})\zeta_{\mathrm{s}}}{\min(\theta v, P_{\mathrm{Sm}})}}^{\infty} f_X(x)f_{V}(v)~\mathrm{d}x~\mathrm{d}v \nonumber \\
& = \underbrace{\int\limits_{v = 0}^{\frac{P_{\mathrm{Sm}}}{\theta}} f_{V}(v)\exp\left(-\frac{(u+N_{0,\mathrm{r}})\zeta_{\mathrm{s}}}{\theta v\lambda_{\mathrm{x}}}\right)\mathrm{d}v}_{T_1} \nonumber \\
&+ \underbrace{\int\limits_{v = \frac{P_{\mathrm{Sm}}}{\theta}}^{\infty} f_{V}(v)\exp\left(-\frac{(u+N_{0,\mathrm{r}})\zeta_{\mathrm{s}}}{P_{\mathrm{Sm}}\lambda_{\mathrm{x}}}\right)\mathrm{d}v}_{T_2},
\label{eq:pout_app1}
\end{align}
where $f_{V}(v)$ is the PDF of the gamma RV $V$. We can write $T_1$ as
\begin{align}
T_1 = \frac{(P_{\mathrm{p}}\lambda_v)^{-L}}{\Gamma(L)}\int\limits_{v = 0}^{\frac{P_{\mathrm{Sm}}}{\theta}}v^{L-1}\exp\left(-\frac{v}{P_{\mathrm{p}}\lambda_v}-\frac{(u+N_{\mathrm{0,r}})\zeta_{\mathrm{s}}}{\theta \lambda_{\mathrm{x}} v}\right)\mathrm{d}v.
\label{eq:T1}
\end{align} 
Scaling the limits of the integral appropriately and using the definition of the incomplete Bessel function as $K_{\nu}(a,b) = \int_{1}^{\infty}\frac{\exp\left(-at - \frac{b}{t}\right)}{t^{\nu+1}} \mathrm{d}t$, we can write \eqref{eq:T1} as \eqref{eq:T1_fin}. 

We can write $T_2$ in \eqref{eq:pout_app1} as
\begin{align}
T_2 &= \frac{\exp\left(-\frac{(u+ N_{0,\mathrm{r}})\zeta_{\mathrm{s}}}{P_{\mathrm{Sm}}\lambda_{\mathrm{x}}}\right)}{\Gamma(L)(P_{\mathrm{p}}\lambda_v)^{L}}\int\limits_{v = \frac{P_{\mathrm{Sm}}}{\theta}}^{\infty} v^{L-1}\exp\left(-\frac{v}{P_{\mathrm{p}}\lambda_v}\right)\mathrm{d}v \nonumber \\
&= \frac{\exp\left(-\frac{(u+ N_{0,\mathrm{r}})\zeta_{\mathrm{s}}}{P_{\mathrm{Sm}}\lambda_{\mathrm{x}}}\right)}{\Gamma(L)} \Gamma\left(L,\frac{P_{\mathrm{Sm}}}{\theta P_{\mathrm{p}} \lambda_v}\right).
\label{eq:T2}
\end{align}

We now derive a closed-form of $\mathbb{P}\left(\gamma_{\mathrm{SD}} \geq \zeta_{\mathrm{s}}|U = u \right)$. We can express $\mathbb{P}\left(\gamma_{\mathrm{SD}} \geq \zeta_{\mathrm{s}}|U = u \right)$ as
\begin{align}
T_3& = \mathbb{P}\left(\gamma_{\mathrm{SD}} \geq \zeta_{\mathrm{s}}|U = u \right) = \mathbb{P}\left(\frac{\min\left(\theta U, P_{\mathrm{Rm}}\right) Y}{Z + N_{0,\mathrm{d}}} \geq \zeta_{\mathrm{s}} \right) \nonumber \\
& = \int_{z = 0}^{\infty} \int\limits_{y = \frac{(z + N_{0,\mathrm{d}})\zeta_{\mathrm{s}}}{\min(\theta u, P_{\mathrm{Rm}})}}^{\infty} f_Y(y) f_Z(z)~\mathrm{d}y~\mathrm{d}z \nonumber \\
& = \frac{\exp\left(-\frac{\zeta_{\mathrm{s}}N_{0,\mathrm{d}}}{\min(\theta u, P_{\mathrm{Rm}})\lambda_{\mathrm{rd}}}\right)}{\Gamma(L)(P_{\mathrm{p}}\lambda_{z})^{L}} \nonumber \\
&\times \int\limits_{z = 0}^{\infty}z^{L-1}\exp\left(-\left(\frac{1}{P_{\mathrm{p}}\lambda_{\mathrm{z}}} + \frac{\zeta_{\mathrm{s}}}{\min(\theta u, P_{\mathrm{Rm}})\lambda_{\mathrm{y}}}\right)z\right)~\mathrm{d}z \nonumber \\
&= \frac{\exp\left(-\frac{\zeta_{\mathrm{s}}N_{0,\mathrm{d}}}{\min(\theta u, P_{\mathrm{Rm}})\lambda_{\mathrm{rd}}}\right)}{(P_{\mathrm{p}}\lambda_{z})^{L}}\left(\frac{1}{P_{\mathrm{p}}\lambda_{\mathrm{z}}} + \frac{\zeta_{\mathrm{s}}}{\min(\theta u, P_{\mathrm{Rm}})\lambda_{\mathrm{y}}}\right)^{-L}\!\!\!\!\!\!.
\label{eq:T3}
\end{align}
Finally, unconditioning on $U$ and using \eqref{eq:T1}, \eqref{eq:T2}, and \eqref{eq:T3}, we obtain the required expression for the secondary outage probability as given in \eqref{eq:ST_EH_out}.

\bibliographystyle{ieeetr}
\bibliography{paper}
\end{document}